\documentclass[reprint,superscriptaddress,aps,pra,floatfix,showpacs]{revtex4-1}

\usepackage{amsmath,amsfonts,amssymb,amsthm,natbib,subfig,epsfig} 
\usepackage{epstopdf}
\usepackage{graphicx}
\usepackage{bm,dcolumn,hyperref}
\hypersetup{pdfstartview={XYZ null null 1.00}} 

\newcommand{\ket}[1]{\left\vert #1 \right\rangle}
\newcommand{\bra}[1]{\left\langle #1 \right\vert}

\newcommand{\ketbra}[3]{\left\vert #1 \right\rangle_{#2} \left\langle #3 \right\vert}
\newcommand{\abs}[1]{\left\vert #1 \right\vert}

\newcommand{\ud}{\mathrm{d}}
\newcommand{\expv}[1]{\left< #1 \right>}

\begin{document}
\title{Enhanced multipartite quantum correlations by non-Gaussian operations}
\author{Ho-Joon Kim}
\affiliation{School of Computational Sciences, Korea Institute for Advanced Study, Hoegiro 87, Dongdaemun, Seoul 130-722, Korea}
\affiliation{Texas A\&M
University at Qatar, Education City, P.O. Box 23874, Doha, Qatar}
\author{Jaewan Kim}
\affiliation{School of Computational Sciences, Korea Institute for Advanced Study, Hoegiro 87, Dongdaemun, Seoul 130-722, Korea}
\author{Hyunchul Nha}
\affiliation{Texas A\&M
University at Qatar, Education City, P.O. Box 23874, Doha, Qatar}
\affiliation{School of Computational Sciences, Korea Institute for Advanced Study, Hoegiro 87, Dongdaemun, Seoul 130-722, Korea}

\begin{abstract}
We study how conditional photon operations can affect multipartite quantum correlations, specifically nonlocality and entanglement, of the continuous variable GHZ states. We find that the violation of the Mermin-Klyshko inequality revealing the multipartite nonlocality can be made stronger with photon subtraction applied on each mode of the original GHZ states, particularly in a weak squeezing regime. Photon addition applied on local modes also turns out to enhance the degree of multipartite nonlocality in a broad range of parameters. We further investigate the effects of the photon operations on the degree of multipartite entanglement by looking into the Gaussian tangle, the fidelity of teleportation network, and the quadrature correlations. We find that photon subtraction applied on two modes enhances those entanglement characteristics in a practical squeezing regime while there is no improvement made by photon addition.
\end{abstract}
\pacs{}
\maketitle

\section{Introduction}
Quantum correlations pose an exotic view of the world that denies the hitherto usual perspective on the world, e.g., local realism. They also provide some new possibilities for information processing in an unprecedented way. Among different forms of quantum correlations, quantum nonlocality has been demonstrated throughout numerous tests \cite{1935einstein,1964bell,1974clauser,1982aspect,1987svetlichny,2002collins,1990mermin,1993klyshko,1998gisin,1998banaszek,2007cavalcanti,2008masanes,2011bancal}, although there still exists a gap of loopholes \cite{1982aspect,1995fry,1998tittel,1998weihs,2001rowe,2008matsukevich,2008cabello,2010ji} to be seamless yet. On a practical side, it is a useful resource for quantum computation, quantum communication, and quantum cryptography \cite{2004nielsen}. Beyond the use of discrete variable systems, quantum informatics utilizing continuous variables (CVs) such as quadrature amplitudes of optical fields has attracted a lot of interest as it is practically easy to manipulate those systems, especially within current technology of quantum optics \cite{2003eisert,2005braunstein}. However, achieving high degree of correlation among CV systems is rather demanding since it requires tremendous energy \cite{1998braunstein} and Gaussian regime that is readily accessible in laboratory is not sufficient to perform universal quantum computation; it needs non-Gaussian operations to be universal \cite{1999lloyd,2005braunstein,2006menicucci,2012weedbrook}.

Typical non-Gaussian operations such as photon subtraction, addition, or superposition of them have been shown to provide some practical advantages for CV quantum information tasks \cite{2008kim}. When the operations succeed, they amplify an optical field with an amount of noise less than that required as the quantum limit \cite{2009ralph,2010marek,2010ferreyrol,2010usuga,2010xiang,2011zavatta,2012kim}. For two mode systems, they can also distill quantum entanglement of bipartite CV systems even for Gaussian states \cite{2009yang,2010takahashi,2010zhang,2011lee,2012navarrete-benlloch} under a noisy environment \cite{2010zhang,Lee2013} beyond the No-Go theorem of Gaussian regime \cite{2002eisert,2002giedke,2002fiurasek} and enhance the violation of local realism \cite{2004olivares,2004nha,2004garcia-patron,2008jeong,2012park}. They can also be useful for quantum communication, e.g., by enhancing the fidelity of the quantum teleportation \cite{2000opatrny,2002cochrane,2003olivares,2007dellanno,2011lee,2012nha}. However, entering into multipartite regime, a study analogous to the bipartite case has been scarce concerning the effect of non-Gaussian operations on multipartite quantum correlations. This is not only because there are few protocols exploiting multipartite nature of a system but also because it is difficult to characterize multipartite correlations in general, especially, for non-Gaussian states in CV quantum information regime \cite{2007adesso}.  Here, extending two-mode systems into multipartite systems, we investigate the effects of photon operations on quantum correlations of a class of multipartite continuous variable systems, the CV GHZ states introduced by van Loock and Braunstein \cite{2000van_loock,2002van_loock,2003aoki}. Quantum nonlocality already has been investigated on this class and the CV GHZ states have shown violations of local realism \cite{2001van_loock,2013lee}. Upon this class, two-mode quantum teleportation protocol naturally extends to multipartite protocol, i.e., teleportation network \cite{2000van_loock} which was also realized in the laboratory \cite{2004yonezawa}. We show that non-Gaussian operations, similar to the results for two-mode systems, can strengthen the multipartite quantum correlations of the CV GHZ states, that is, enhance the violation of local realism, the Gaussian tangle, quadrature correlations, and the fidelity of the teleportation network.

This paper is organized as follows. In Sec.~\ref{sec:formalism} we recapitulate the phase space formalism that will be used in the following sections and also introduce the non-Gaussian operations, photon subtraction and addition, described within that formalism. In Sec.~\ref{sec:entanglement} we investigate the Gaussian tangle of the CV GHZ states with photon operations. In Sec.~\ref{sec:nonlocality}, we then investigate the quantum nonlocality of the non-Gaussian states produced with those photon operations including detection efficiency in our analysis. In Sec.~\ref{sec:teleportation}, we study a quantum communication task specifically the fidelity of the teleportation network together with the multipartite EPR (quadrature) correlations. In Sec.~\ref{sec:conclusion}, we summarize our main results.

\section{CONTINUOUS VARIABLE GHZ STATES WITH PHOTON OPERATIONS}\label{sec:formalism}
\subsection{PHASE SPACE FORMALISM AND THE CONTINUOUS VARIABLE GHZ STATES}
An $N$-mode CV state can be described within phase space formalism. Denoting the quadrature operators collectively as
\begin{equation}
X=(x_1,p_1,x_2,p_2,\cdots,x_N,p_N)^T,
\end{equation}
where $x_i=(a_i+a_i^\dagger)/\sqrt{2}$, $p_i=(a_i-a_i^\dagger)/(i\sqrt{2})$ are the quadrature operators for the $i$-th mode, the characteristic function for a state $\rho$ is defined as $\chi(\eta)\equiv \mathrm{Tr}\left[\rho \exp(i \eta^T \cdot X)\right]\quad (\eta \in\mathbb{R}^{2N})$ and the Wigner distribution is defined as its Fourier transform, $W(\xi)\equiv \int \frac{\ud^{2N} \eta}{(2\pi)^{2N}}\chi(\eta)\exp(-i \xi^T \cdot \eta) \quad (\xi \in\mathbb{R}^{2N})$. An $N$-mode Gaussian state is a state having a Gaussian Wigner distribution
\begin{equation}
W(\xi)=\frac{1}{\sqrt{\mathrm{det}(2\pi V)}}\exp\left[-\frac{1}{2}(\xi-\mu)^T V^{-1}(\xi-\mu)\right],
\end{equation}
where $\mu=\mathrm{Tr}(\rho X)\in\mathbb{R}^{2N}$ is the mean vector of the quadratures and $V$ is the covariance matrix defined as
\begin{equation}
V_{kl}=\expv{\{\Delta X_k,\Delta X_l\}_s}=\frac{1}{2}\expv{X_k X_l+X_l X_k}-\expv{X_k}\expv{X_l}
\end{equation}
with $\Delta X_k\equiv X_k-\expv{X_k}$. 

A CV GHZ state is a class of genuinely multipartite entangled Gaussian states, which corresponds to the continuous version of the GHZ states for qudits \cite{2000van_loock}. It can be generated by injecting a $p$-squeezed vacuum mode with a squeezing parameter $r_1$ and $N-1$ $x$-squeezed vacuum modes with a squeezing parameter $r_2$, respectively, through a series of beam splitters. A CV GHZ state is described by its mean vector $\mu=0$ and its covariance matrix 
\begin{equation}
V^{(N)}=\begin{pmatrix}
a&0&c&0&c&0&\cdots\\
0&b&0&d&0&d&\cdots\\
c&0&a&0&c&0&\cdots\\
0&d&0&b&0&d&\cdots\\
c&0&c&0&a&0&\cdots\\
0&d&0&d&0&b&\cdots\\
\vdots&\vdots&\vdots&\vdots&\vdots&\vdots&\vdots
\end{pmatrix},
\end{equation}
where
\[
\begin{array}{ll}
a=\frac{1}{2N}e^{2r_1}+\frac{N-1}{2N}e^{-2r_2},&b=\frac{1}{2N}e^{-2r_1}+\frac{N-1}{2N}e^{2r_2},\\
c=\frac{1}{2N}(e^{2r_1}-e^{-2r_2}),&d=\frac{1}{2N}(e^{-2r_1}-e^{2r_2}).
\end{array}
\]
For infinite squeezing parameters $r_1,\,r_2\rightarrow \infty$, the quadratures of the state are correlated as
\begin{equation}\label{eq:quadrature correlations}
\langle[\Delta (x_i-x_j)]^2\rangle\rightarrow 0,\,\langle(\Delta\sum_i p_i)^2\rangle\rightarrow 0.
\end{equation}

Note that the covariance matrix is invariant under a permutation of modes and there is no intermode or intramode $x$-$p$ correlations. One can adjust the two squeezing parameters so that the variances of the quadratures $x$ and $p$ of each mode can be made equal, called `unbiased'; in general the two variances may be different, called  `biased.' In the following, we investigate biased three-mode CV GHZ states with a squeezing parameter $r_1=r_2=r$, as it may be natural to produce in laboratory.

\subsection{PHOTON SUBTRACTION AND ADDITION}
Photon subtraction on a mode can be implemented by mixing the mode with an ancillary vacuum mode via a near-transparent beam splitter ($t\lesssim 1$) and then, postselecting the `on' event at an on-off detector on the ancillary mode. For example, let $V_0$ denote the covariance matrix of an initial CV GHZ state of three modes $A$, $B$, and $C$. When a photon subtraction is performed on the mode $A$, the beam splitter operation with transmittivity $t$ ($|t|^2+|r|^2=1$) is described by a symplectic transformation $S_{BS}$ acting on the covariance matrix of the total state including an ancillary vacuum mode $D$, that is,
\begin{equation}
V=S_{BS}\left (V_0\oplus \frac12 I_2\right )S_{BS}^T\equiv\begin{pmatrix}
\Gamma&M\\
M^T&\Delta
\end{pmatrix}
\label{subtraction_formalism}
\end{equation}
with
\begin{equation}
S_{BS}=\begin{pmatrix}
tI_2&0&-rI_2\\
0&I_4&0\\
rI_2&0&tI_2
\end{pmatrix}.
\end{equation}
Here $\Gamma$ and $\Delta$ are $6\times6$ and $2\times2$ covariance matrices of the three modes and the ancillary mode, respectively, after applying the beam splitter operation. On the other hand, $6\times2$ matrix $M$ represents correlations between the three modes and the ancillary mode, and $I_d$ is the identity matrix of dimension $d$.
The `on' event at the detector on the mode $D$ is described by a projection operator $\Pi_D=I-\ketbra{0}{D}{0}$ on the ancillary mode $D$. Denoting the total state including the ancillary mode $D$ after the beam splitter operation as $\rho'$, the state after post-selection by the detector is described as $\rho_f=\mathrm{Tr}_D(\rho' \Pi_D)$ (without normalization). Using the Wigner distribution of the projection operator $\Pi_D$,
\begin{equation}
W_{\rm on}(\zeta\in \mathbb{R}^2)=\frac{1}{2 \pi}-\frac{1}{\pi}e^{-\zeta_1^2-\zeta_2^2},
\end{equation}
the final state $\rho_f$ is written as (without normalization)
\begin{eqnarray}
\tilde{W}_\rho(\xi\in \mathbb{R}^{6})&=&2\pi \int \mathrm{d}^2 \zeta\, W_{\rho'}(\xi,\zeta)\,W_{\rm on}(\zeta)\nonumber \\
&=&\mathcal{N}(\xi;0,\Gamma) \nonumber \\
&&-\frac{\mathcal{N}\left(\xi;0,\Gamma-M(\Delta+I_2/2)^{-1}M^T\right)}{\sqrt{\mathrm{det}(\Delta+I_2/2)}}, \label{wigner_distribution_photon_subtraction}
\end{eqnarray}
where the notation $\mathcal{N}(\xi;\mu,C)$ means a multivariate normal distribution with mean vector $\mu$ and covariance matrix $C$. Photon subtraction on other modes can be similarly described.

Photon addition can also be treated in a similar way to photon subtraction, by replacing the beam splitter with a nondegenerate parametric amplifier having weak interaction strength ($s\ll 1$). For instance, among three modes A, B, and C, a nondegenerate parametric amplifier on mode $A$ with an ancillary mode $D$ is described by Eq.~\eqref{subtraction_formalism} with a symplectic transformation
\begin{equation}
S_{ND}=\begin{pmatrix}
I_2\cosh{s}&0&\sigma_z \sinh{s}\\
0&I_4&0\\
\sigma_z \sinh{s}&0&I_2 \cosh{s}
\end{pmatrix},
\end{equation}
where $\sigma_z=\begin{pmatrix}
1&0\\0&-1
\end{pmatrix}$ is one of the Pauli operators.

In the following sections, we investigate Gaussian multipartite entanglement (tangle), multipartite nonlocality, the fidelity of the teleportation network, and the multipartite EPR correlations of the three-mode CV GHZ states for which the photon operations are applied on one mode, two modes or three modes. We use beam splitters with transmittivity $t=0.99$ for photon subtraction and nondegenerate parametric amplifiers with interaction strength $s=0.01$ throughout the article, which are currently available in experiments \cite{2008zavatta}.

\section{MULTIPARTITE ENTANGLEMENT}\label{sec:entanglement}
First we investigate multipartite entanglement of the three-mode CV GHZ state with photon operations. To the lowest orders of the squeezing parameter $r$, a three-mode biased CV GHZ state in the Fock basis turns out to be
\begin{widetext}
\begin{eqnarray}
\ket{\rm GHZ}&=&\exp\left [-\frac{r}{6}(a^{\dag 2}+b^{\dag 2}+c^{\dag 2}-4a^{\dag}b^{\dag}-4b^{\dag}c^{\dag}-4c^{\dag}a^{\dag}-a^2-b^2-c^2+4a b+4b c+4c a)\right ]\ket{0}\\
&\approx& \ket{0}-\frac{\sqrt{2}}{6}r(\ket{200}+\ket{020}+\ket{002})+\frac{2}{3}r(\ket{110}+\ket{011}+\ket{101})+O(r^2).\label{eq:approximate GHZ}
\end{eqnarray}
\end{widetext}
It then follows that the CV GHZ states with photon operations are given by (without normalization constants)
\begin{eqnarray}
a\ket{\rm GHZ}&\propto & -\frac{1}{3}r\ket{100}+\frac{2}{3}r(\ket{010}+\ket{001})\nonumber\\
&&+O(r^2)\label{eq:approximate subA}\\
ab\ket{\rm GHZ}&\propto &\frac{2}{3}r\ket{000}+O(r^2)\label{eq:approximate subAB}\\
abc\ket{\rm GHZ}&\propto &\frac{3}{4}r^2(\ket{100}+\ket{010}+\ket{001})\nonumber\\
&&+O(r^3)\label{eq:approximate subABC}\\
a^\dag\ket{\rm GHZ}&\propto &\ket{100}-\frac{\sqrt{2}}{6}r(\sqrt{3}\ket{300}+\ket{120}+\ket{102})\nonumber\\
&&+\frac{2}{3}r(\sqrt{2}\ket{210}+\ket{111}+\sqrt{2}\ket{201})\nonumber\\
&&+O(r^2)\label{eq:approximate addA}\\
a^\dag b^\dag\ket{\rm GHZ}&\propto &\ket{110}-\frac{\sqrt{2}}{6}r(\sqrt{3}\ket{310}+\sqrt{3}\ket{130}+\ket{112})\nonumber\\
&&+\frac{2\sqrt{2}}{3}r(\sqrt{2}\ket{220}+\ket{121}+\ket{211})\nonumber\\
&&+O(r^2)\label{eq:approximate addAB}\\
a^\dag b^\dag c^\dag\ket{\rm GHZ}&\propto &\ket{111}-\frac{1}{\sqrt{6}}r(\ket{311}+\ket{131}+\ket{113})\nonumber\\
&&+\frac{4}{3}r(\ket{122}+\ket{212}+\ket{221})\nonumber\\
&&+O(r^2).\label{eq:approximate addABC}
\end{eqnarray}
Note that the CV GHZ state with photon subtraction on all three modes maintains tripartite entanglement even in the limit of vanishing squeezing $r\rightarrow0$, which actually becomes a $W$-type entangled state [Eq.\eqref{eq:approximate subABC}]. The CV GHZ state with photon subtraction on one mode also possesses some entanglement in the same limit [Eq.\eqref{eq:approximate subA}], while other states lose quantum correlations with $r$ decreasing.

When the squeezing parameter $r$ is not very small, each mode of all considered states  resides in infinite dimensional space, thus a suitable measure of multipartite entanglement is not readily available. W here study the Gaussian tangle (or the Gaussian residual entanglement) that corresponds to the CV version of the tangle for qubits \cite{2006adesso}. The tangle is a quantity defined from an entanglement monogamy in terms of squared concurrences \cite{2000coffman}. For a pure tripartite state $\rho_{ABC}$, the Gaussian tangle is defined as
\begin{equation}
E_{\tau}(\rho_{ABC})=\min_{(i,j,k)}E_N^{i|jk}(\rho_{ABC})-E_N^{i|j}(\rho_{ij})-E_N^{i|k}(\rho_{ik}),\label{eq:gaussian tangle def}
\end{equation}
where $(i,j,k)\in \rm Sym \{A,B,C\}$ and $E_N^{i|j}(\rho_{ij})=(-\log_2 \Vert \rho_{ij}^{T_i}\Vert_1)^2$ is the square of the logarithmic negativity of the partition $i|j$ of $\rho_{ij}$ \cite{2002vidal}. Since the Gaussian tangle is an entanglement monotone under Gaussian local operations and classical communications for a pure three-mode Gaussian state, we may consider it as a witness to multipartite entanglement of Gaussian characteristic. Note that, due to its definition, the Gaussian tangle of a given CV GHZ state with photon operations does not depend on which modes undergo the photon operations. The tangle depends only on the number of modes where the photon operations are applied and it turns out that the bipartitions for the Gaussian tangle in Eq.~\eqref{eq:gaussian tangle def} are those in which a mode undergoing the photon operation is referred to as $i$.

\begin{figure}[htb]
\centering
\includegraphics[width=0.4\textwidth]{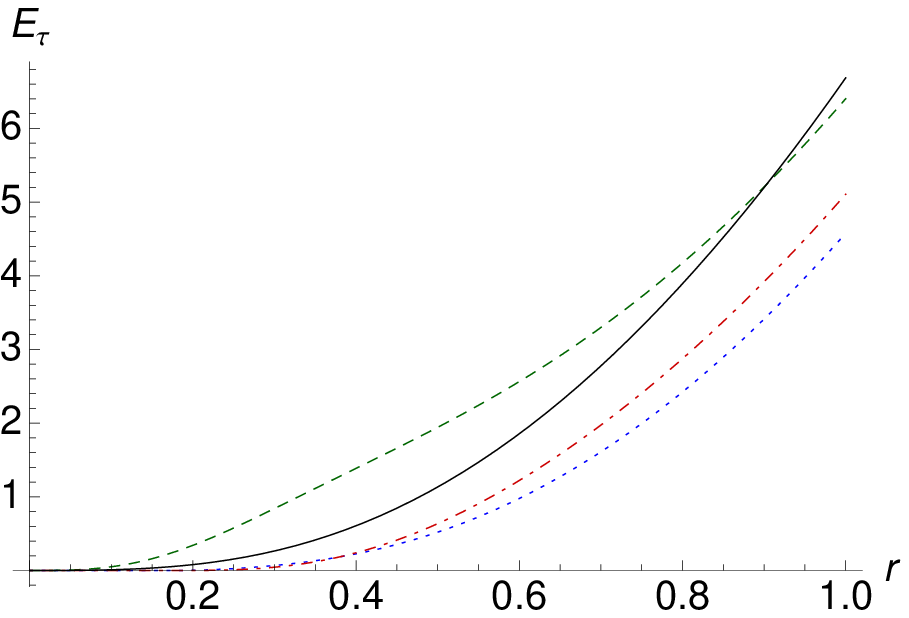}
\includegraphics[width=0.4\textwidth]{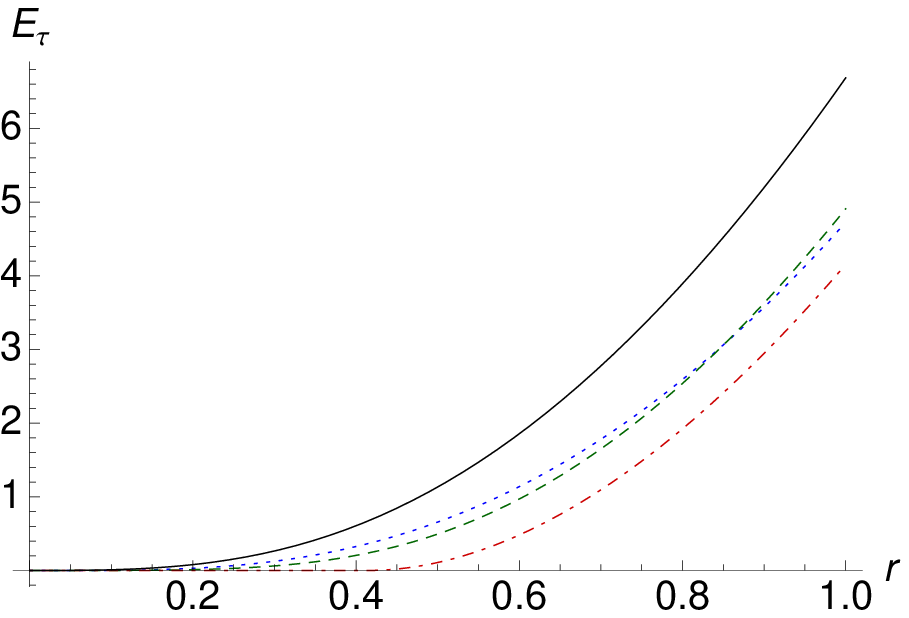}
\caption{(Color online) : Gaussian tangle as a function of the squeezing parameter $r$: (Top) photon subtraction (bottom) photon addition, on mode A (blue, dotted), on modes A and B (green, dashed), on modes A, B, and C (red, dot-dashed) in comparison to the CV GHZ state (black solid). See main text.}
\label{fig_Gaussian_tangle}
\end{figure}
Fig.~\ref{fig_Gaussian_tangle} shows that photon subtraction enhances the Gaussian tangle when it is applied on two modes of the CV GHZ state in a weak squeezing regime, while other cases do not have such an effect. For example, as seen from Eqs.\eqref{eq:approximate subABC} and \eqref{eq:approximate subABC}, photon subtractions on one mode or three modes result in a certain non-Gaussian multipartite entanglement in the weak squeezing limit $r\to 0$, however they do not enhance the multipartite entanglement of Gaussian nature. Moreover, photon additions do not enhance the Gaussian tangle of the CV GHZ state at all.

\section{MULTIPARTITE NONLOCALITY}\label{sec:nonlocality}
There have been various tests proposed to reveal multipartite nonlocality of quantum states and here we use the Mermin-Klyshko (MK) inequality \cite{1990mermin,1993klyshko,1998gisin} to investigate the effects of non-Gaussian operations on the degree of multipartite nonlocality. This inequality can be employed for CV systems using displaced parity operators as the dichotomic variables \citep{1998banaszek}.

Let $a_k$ and $a_k'$ be two observables of mode $k$ having outcomes $\pm 1$. Then the MK polynomial is defined recursively as
\begin{eqnarray}
B_2&=&a_1 (a_2+a_2')+a_1' (a_2-a_2'),\\
B_n&=&\frac{1}{2} (a_n+a_n') B_{n-1}+\frac{1}{2} (a_n-a_n') B_{n-1}',
\end{eqnarray}
where $B_n'$ can be obtained from $B_n$ by exchanging $a_k$ and $a_k'$. Within local realistic theories, we have the MK inequality which gives the upper bound on the expectation value of the MK polynomial such that $|\expv{B_n}|\leq2$; Any violation of the MK inequality witnesses multipartite quantum nonlocality. Quantum mechanics could allow the expectation value of the MK polynomial up to $2^{(n+1)/2}$ that grows with $n$ \cite{1998gisin}.

The MK inequality can be tested for a CV multipartite state with the displaced parity operators as the dichotomic variables \cite{1998banaszek}. Using the fact that the Wigner distribution corresponds to the expectation value of the displaced parity operators as
\begin{equation}
W(\alpha_1,\alpha_2,\alpha_3)=\frac{1}{\pi^3}\expv{\pi_1(\alpha_1)\pi_2(\alpha_2)\pi_3(\alpha_3)},
\end{equation}
where $\pi_i(\alpha_i)=D_i(\alpha_i)\pi_i D_i^\dag(\alpha_i)$ are the displaced parity operators with the parity operator $\pi_i=(-1)^{n_i}$ \cite{1977royer}, the MK polynomial for a three mode state is written as
\begin{eqnarray}
B_3&=&\pi^3\left\{ W(\alpha_1,\alpha_2,\alpha_3')+W(\alpha_1,\alpha_2',\alpha_3)\right.\nonumber \\
&&\left.+W(\alpha_1',\alpha_2,\alpha_3)-W(\alpha_1',\alpha_2',\alpha_3') \right\},
\end{eqnarray}
where $\alpha_i$ and $\alpha_i'$ are two different displacement parameters for the two observable settings of the $i$-th mode. For comparison to the previous work \cite{2001van_loock}, we use the setting $\alpha_i=0$ and $\alpha_i'=i x$ ($x\in \mathbb{R}$), which has shown the violation of the inequality $|B_3|\leq 2$ up to $|B_3|\approx 2.324$ for the CV GHZ states. In the following, we probe the effects of photon operations on the multipartite nonlocality of the CV GHZ states.

\begin{figure}[htb]
\includegraphics[width=0.4\textwidth]{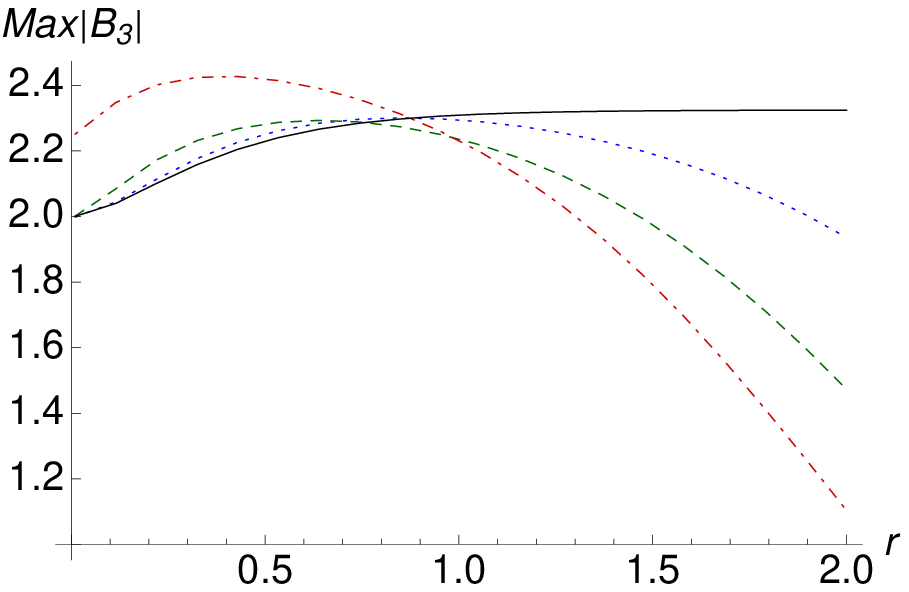}
\includegraphics[width=0.4\textwidth]{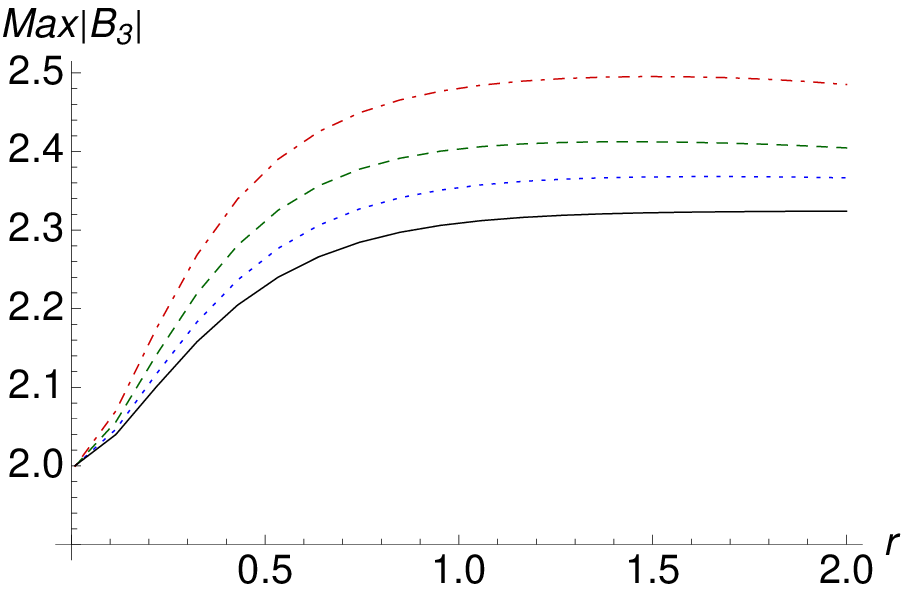}
\caption{(Color online) Expectation value of the MK polynomial as a function of the squeezing parameter $r$: (Top) photon subtraction (bottom) photon addition, on mode A (blue, dotted), on modes A and B (green, dashed), on modes A, B, and C (red, dot-dashed) in comparison to the CV GHZ state (black solid). Each point of the graphs is maximized with respect to the magnitude $\abs{x}$ of the displaced parity operator. See main text.}
\label{fig_MK_inequality}
\end{figure}
In Fig.~\ref{fig_MK_inequality}, we show the MK polynomial values maximized with respect to $x$ as a function of the squeezing parameter $r$. The top figure of Fig.~\ref{fig_MK_inequality} shows that, compared to the case of the CV GHZ state, photon subtraction enhances the violation of the MK inequalities particularly in a small squeezing regime. The maximum values of MK polynomial of each scheme turn out to be $\rm Max \abs{B_3}\approx 2.301,2.293,2.428$ for one-photon, two-photon, and three-photon subtracted CV GHZ states, respectively. Thus photon subtraction on all three-modes gives a larger maximum value than that of the CV GHZ state. Remarkably, this improved maximum is achieved for a smaller squeezing parameter $r$. The degree of violation diminishes as the squeezing increases for all the cases and the case of photon subtraction on all three modes goes down most steeply.

On the other hand, bottom figure of Fig.~\ref{fig_MK_inequality} shows that photon addition generally enhances the violation of the MK inequality over a large range of the squeezing parameter $r$. The enhancement clearly increases with the number of modes on which the operation is applied and the maximum values of the MK polynomial are $\rm Max \abs{B_3}\approx 2.368,2.412,2.495$ for one-mode, two-mode, and three-mode photon added CV GHZ states, respectively. Obviously, the maximum value of the MK polynomial $\rm Max \abs{B_3}\approx 2.324$ of the CV GHZ state are surpassed by the CV GHZ states with non-Gaussian operations. Moreover, the maximum values of those non-Gaussian states are achieved in a practically less demanding regime as no large squeezing parameter $r$ is needed to achieve a high value of $\abs{B_3}$.

We also investigate the effect of the detection efficiency on the multipartite nonlocality of the CV GHZ states with photon operations. Detection loss on each mode can be modeled as an interaction with a vacuum mode via a beam splitter with transmittivity $t$ which corresponds to the detection efficiency $\eta=\abs{t}^2$ \cite{2005leonhardt}. When each mode of an arbitrary $n$-mode Gaussian state with zero mean vector and covariance matrix $V$ experiences such an interaction with a vacuum mode through a beam splitter with transmissivity $t$, a simple calculation shows that the resulting state becomes a Gaussian state with covariance matrix $V'$
\begin{equation}
V'=\abs{t}^2 V+\frac{\abs{r}^2}{2}I_{2n}.
\end{equation}
Since the CV GHZ state with photon operations is a statistical sum of Gaussian states as in Eq.~\eqref{wigner_distribution_photon_subtraction}, the above formula enables us to calculate the final state with detection loss.

\begin{figure}[ht]
\centering
\includegraphics[width=0.4\textwidth]{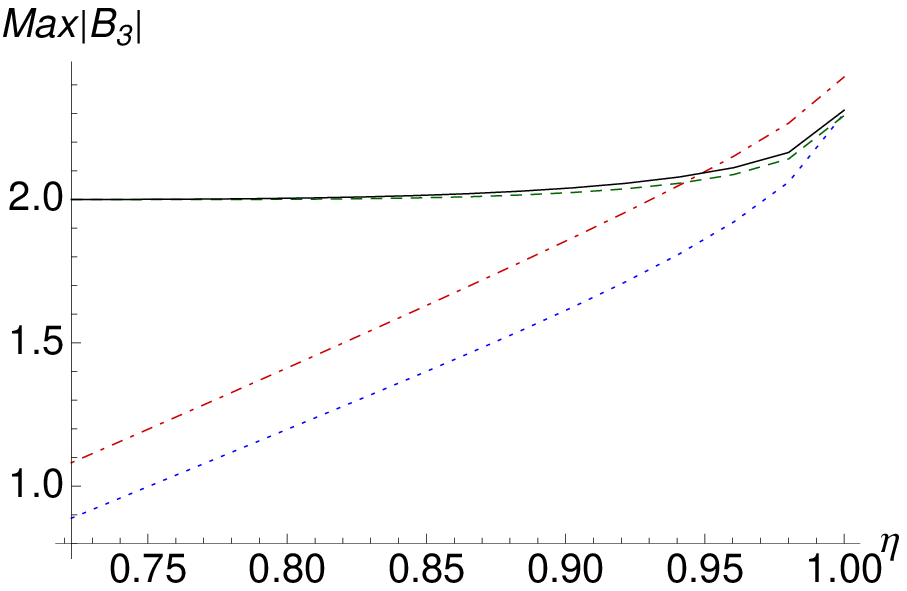}
\includegraphics[width=0.4\textwidth]{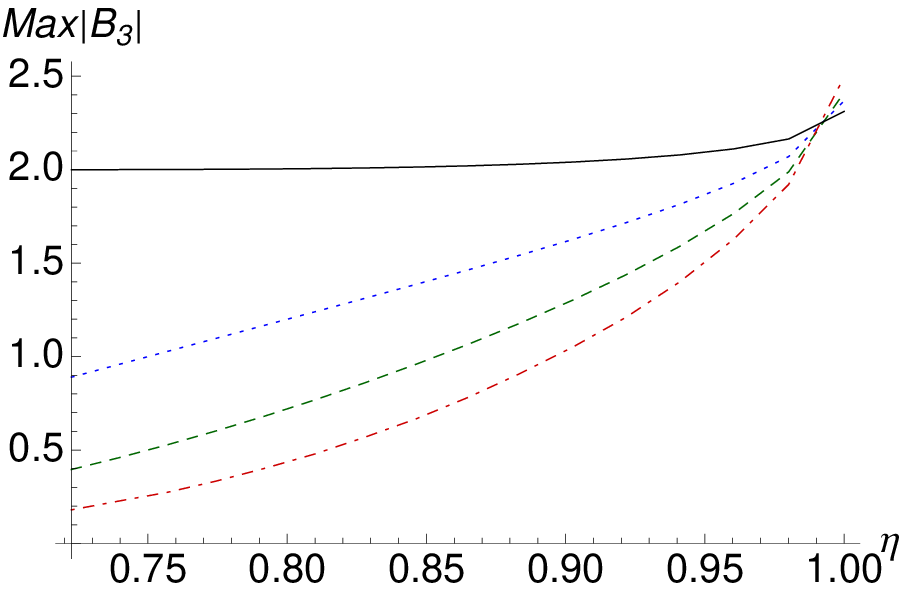}
\caption{(Color online) Expectation value of the MK polynomial against detection efficiency $\eta$. (Top) photon subtraction with noise (bottom) photon addition with noise, on mode A (blue, dotted), on modes A and B (green, dashed), on modes A, B, and C (red, dot-dashed) in comparison to the CV GHZ state (black solid). Each point of the graphs is maximized with respect to both the magnitude $\abs{x}$ of the displaced parity operator and the squeezing parameter $r\in (0,2)$. See main text.}
\label{fig_MK_inequality_noise}
\end{figure}
Fig.~\ref{fig_MK_inequality_noise} shows the multipartite nonlocality of the CV GHZ state and those with photon operations, maximized with respect to both the magnitude $\abs{x}$ of the displaced parity operator and the squeezing parameter $r\in (0,2)$ within practical reach. We see that the original CV GHZ state shows the violation of MK inequality over a larger range of $\eta$ than other schemes. However, the multipartite nonlocalities of the CV GHZ states with either photon subtraction or photon addition on all three modes (red dot-dashed curves) are overall stronger for a high detection efficiency $\eta$ than that of the original GHZ state (black solid curves). The CV GHZ state with photon subtraction on two modes shows similar degree of nonlocality to that of the original CV GHZ state. The threshold detection efficiency for the original CV GHZ state is $\eta\approx 0.694$, and those for the CV GHZ state with photon subtraction on one mode, two modes, and three modes are $\eta\approx 0.972,0.750,0.931$, respectively. The CV GHZ states with photon addition have still higher nonlocality near ideal detection efficiency, although they are more fragile against detection loss than the CV GHZ states or those with photon subtraction. The threshold detection efficiencies of the CV GHZ states with photon addition on one mode, two modes, and three modes are $\eta\approx 0.972,0.982,0.986$, respectively.

\section{FIDELITY OF THE TELEPORTATION NETWORK PROTOCOL}\label{sec:teleportation}
One application of multipartite entangled states for quantum communication is the teleportation network protocol, which is an extension of the Braunstein-Kimble scheme for two-mode CV quantum teleportation \cite{1998braunstein}. In the teleportation network protocol, a sender possessing one mode of a multipartite entangled state can transfer an unknown input state to a receiver possessing another mode of the source, by the help of the remaining parties with the feed-forward of their measurement results to the receiver \cite{2000van_loock}. We examine a three mode scheme where Alice, Bob, and Charlie share one mode of a three-mode entangled state $\rho_{\rm ent}$ each, and Alice wants to send an unknown input state $\rho_{\rm in}$ to Charlie. As in the two-mode teleportation scheme, Alice first mixes her mode with the input state through a 50:50 beam splitter and measures the $x$-quadrature of the output mode $u$ and the $p$-quadrature of the other output mode $v$ given by
\begin{equation}
x_u=\frac{1}{\sqrt{2}}(x_{\text{in}}-x_1),\, p_v=\frac{1}{\sqrt{2}}(p_{\text{in}}+p_1).
\end{equation}
She sends her measurement results $x_u$ and $p_v$ to Charlie. Bob also measures the $p$-quadrature of his mode and sends the result $p_2$ to Charlie. Charlie obtains the output state after displacing his mode by the amount of $\sqrt{2} (x_u+ip_v)+i g p_2$, where $g$ is a gain factor that can be adjusted to give an optimal fidelity. Without the information on $p_2$, the output fidelity may go below the classical bound as the squeezing parameter $r$ increases \cite{2002van_loock}. A high fidelity that can be achieved only with other remaining party's help can be regarded as an evidence of multipartite entanglement. Expressing the multipartite entanglement source and the input state in the Wigner distribution $W_{\rm ent}(\xi_1,\xi_2,\xi_3)$ and $W_{\rm in}(\xi_0)$ $(\xi_i=(x_{i},p_{i})\in \mathbb{R}^2)$, respectively, the Wigner distribution of the output state is given by \cite{2001van_loock_b}
\begin{widetext}
\begin{equation}
\overline{W}_{\mathrm{out}}(x_3,p_3)=\int \mathrm{d}^2\xi_0\mathrm{d}^2\xi_1\mathrm{d}^2\xi_2\, W_{\mathrm{in}}(\xi_0)W_{\mathrm{ent}}(\xi_1,\xi_2,x_{3}-(x_{0}-x_{1}),p_{3}-(p_{0}+p_{1})-g p_{2}).
\end{equation}
\end{widetext}
If we use a Gaussian state with mean vector $\mu$ and covariance matrix $V$ as an entanglement source and a coherent state $\ket{\alpha}$ as an input state, the output fidelity turns out to be
\begin{eqnarray}
F_{\text{avg}}&=&\bra{\alpha}\overline{\rho}_{\text{out}}\ket{\alpha}\nonumber\\
&=&2\pi \int \mathrm{d}^2 \xi\, W_{\alpha}(\xi)\overline{W}_{\mathrm{out}}(\xi)\nonumber\\
&=&4\{\det V \det(L_1+K^TL_2K)\}^{-1/2}
\label{output_fidelity}
\end{eqnarray}
with 
\begin{eqnarray*}
&&L_1=\begin{pmatrix}
2I_2& & \\&0_{4\times 4}& \\ & &2I_2
\end{pmatrix},\,
L_2=\begin{pmatrix}
0_{2\times 2}&&\\&&V^{-1}
\end{pmatrix},\\
&&K=\begin{pmatrix}
&0_{2\times 8}&\\
0_{4\times 2}&I_4&0_{4\times 2}\\
\begin{matrix}
-1&0\\0&-1
\end{matrix}&\begin{matrix}
1&0&0&0\\0&-1&0&-g
\end{matrix}&\begin{matrix}
1&0\\0&1
\end{matrix}
\end{pmatrix},
\end{eqnarray*}
where $0_{m\times n}$ is a $m\times n$ matrix whose elements are all zero.
Should we have used the CV GHZ state with infinite squeezing as the multipartite entanglement source, the output state would become identical to the input state, but not for a finite squeezing. Note that the fidelity in Eq.~\eqref{output_fidelity} is independent of the amplitude $\alpha$ of the input coherent state.

\begin{figure}
\includegraphics[width=0.4\textwidth]{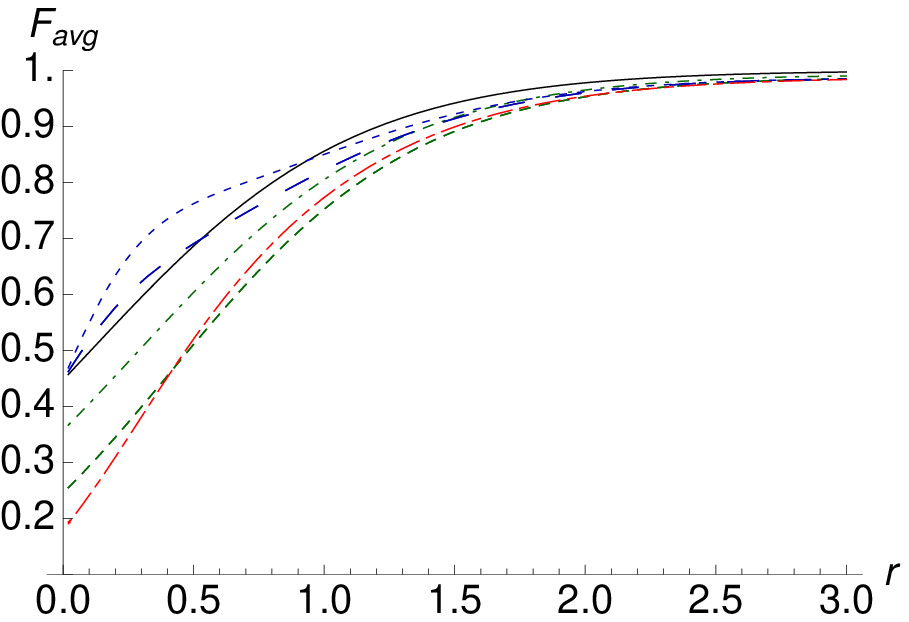}
\includegraphics[width=0.4\textwidth]{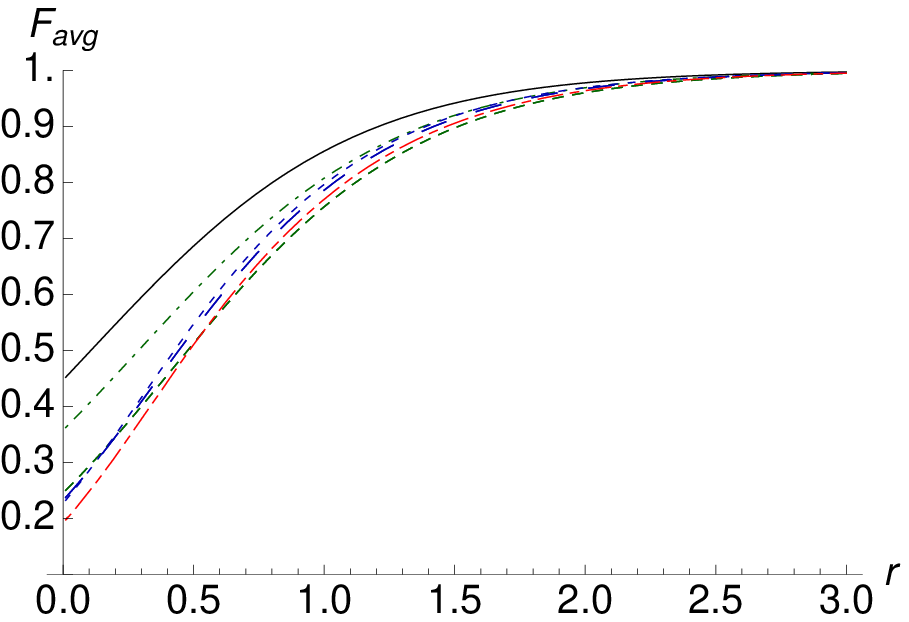}
\caption{(Color online) Fidelity of the teleportation network protocol with unit gain $g=1$ using the CV GHZ states with photon operations as a function of the squeezing parameter $r$. Top curves for photon subtraction and bottom curves for photon addition on mode A (green, small-dashed), mode B (green, dot-small-dashed), mode C (green, small-dashed), mode A and B (blue, long-dashed), mode A and C (blue, dotted), mode B and C (blue, long-dashed) and mode A, B and C (red, dot-long-dashed) in comparison to the CV GHZ state (black solid). The curves for photon operation on mode A and those on mode C coincide. The curves for photon operations on both modes A and B and those on modes B and C also coincide.}
\label{fidelity_unit_gain}
\end{figure}

\begin{figure}
\includegraphics[width=0.4\textwidth]{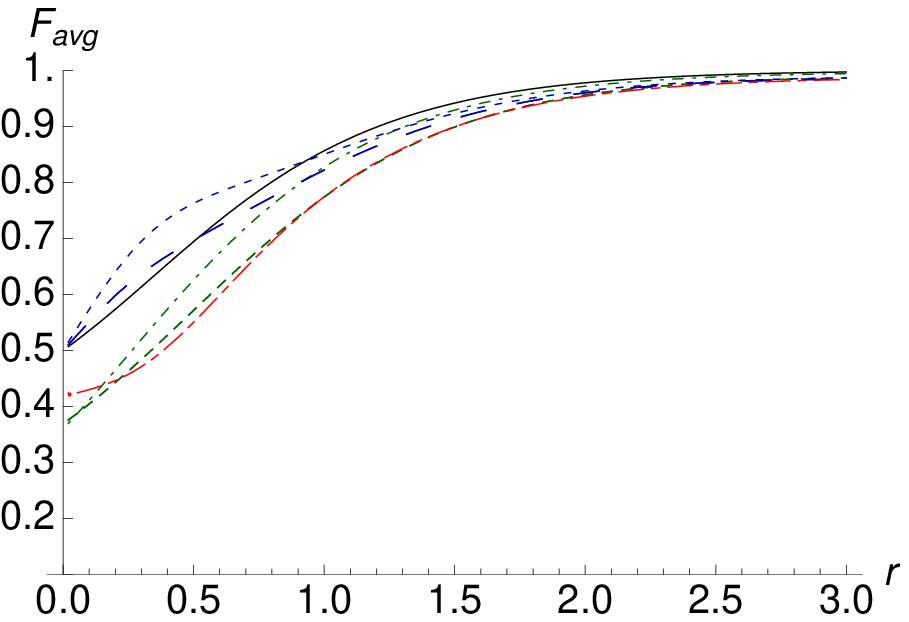}
\includegraphics[width=0.4\textwidth]{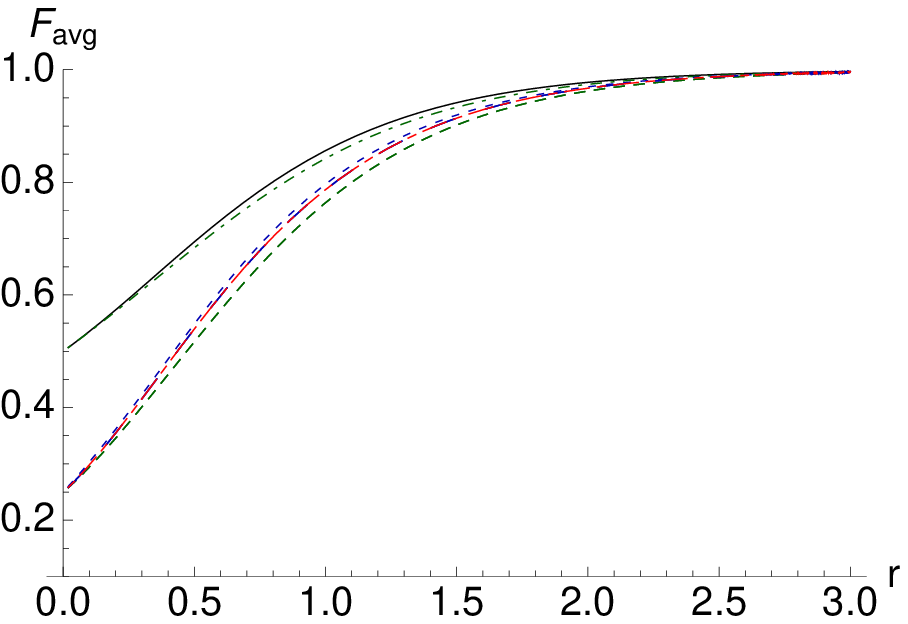}
\caption{(Color online) Fidelity of the teleportation network protocol maximized over the gain $g$ using the CV GHZ states with photon operations as a function of the squeezing parameter $r$. Top curves for photon subtraction and bottom curves for photon addition on mode A (green, small-dashed), mode B (green, dot-small-dashed), mode C (green, small-dashed), mode A and B (blue, long-dashed), mode A and C (blue, dotted), mode B and C (blue, long-dashed) and mode A, B and C (red, dot-long-dashed) in comparison to the CV GHZ state (black solid). The curves for photon operation on mode A and those on mode C coincide. The curves for photon operations on modes A and B and those on modes B and C also coincide.}
\label{fidelity_optimal_gain}
\end{figure}

The Wigner distributions of the CV GHZ states with photon operations applied can be written as a sum of Wigner distributions of Gaussian states with zero mean, for instance, Eq.~\eqref{wigner_distribution_photon_subtraction}. Therefore, using the CV GHZ states with photon operations as the multipartite entanglement source, we can calculate the fidelity of the teleportation network using  Eq.~\eqref{output_fidelity} for each Gaussian Wigner distribution in the summation. Figs.~\ref{fidelity_unit_gain} and \ref{fidelity_optimal_gain} show the fidelity of the unit gain scheme ($g=1$) and the fidelity maximized over $g$, respectively, of the teleportation network protocol using the CV GHZ states with photon operations as a function of the squeezing parameter $r$. Note that unlike the case of nonlocality test, each party plays a different role in the teleportation network protocol so that the output fidelity depends on the particular modes photon operations are applied on.


\subsection{UNIT GAIN: $g=1$}
The top figure of Fig.~\ref{fidelity_unit_gain} shows that photon subtraction enhances the fidelity of the teleportation network protocol in a weak squeezing regime when it is applied on two modes of the CV GHZ state; photon subtraction on modes $A$ (sender) and $C$ (receiver) gives the best enhancement. The fidelities of other cases, photon subtraction on modes $A$ and $B$ and on modes $B$ and $C$, coincide and show a less enhancement. On the other hand, neither photon subtraction on one mode nor photon subtraction on all three modes improves the fidelity. Photon subtraction on mode B (helper) gives the best fidelity among one-mode subtraction schemes and fidelities of both cases, photon subtraction on mode A and on mode C, coincide with reduced fidelity. Different from the case of the two-mode teleportation scheme using two-mode squeezed states with the photon operations \cite{2000opatrny}, photon subtraction on all three modes involved does not give the best result in the teleportation network protocol. 

The bottom figure of Fig.~\ref{fidelity_unit_gain} shows that photon addition is unable to improve the fidelity wherever the operation is applied. Let alone enhancement, among  photon addition on one mode schemes, photon addition applied on mode B (helper) gives the best fidelity. Photon addition on three modes shows the worst result in a weak squeezing regime. Within two-mode photon addition schemes, choosing mode A (sender) and mode C (receiver) shows a slightly better fidelity than other cases in a weak squeezing regime and the remaining photon addition schemes applied on two modes give the same fidelity.

Note that the CV GHZ state, when there is no entanglement ($r=0$), gives a fidelity $F_{\rm avg}=1/\sqrt{5}$ worse than the best fidelity $F_{\rm classical}=1/2$ of classical teleportation \cite{2000braunstein}. This is because an additional vacuum noise from Bob's measurement result deteriorates the output state. However, as the correlation in the source becomes stronger, it reaches the classical best fidelity 1/2 around $r\approx 0.107$ and approaches the ideal fidelity $F_{\rm avg}=1$ as the squeezing parameter $r$ further grows. As for photon subtracted CV GHZ states, the threshold squeezing parameters to overcome the best fidelity 1/2 of classical means are $r_A=r_C\approx 0.481$, $r_B\approx 0.291$, $r_{AB}=r_{BC}\approx 0.080$, $r_{AC}\approx 0.060$, and $r_{ABC}\approx 0.469$, where subscripts denotes the modes on which photon subtraction is applied. Regarding an achievable squeezing parameter in practice, $r\approx 0.876$ corresponding to $7.6\, \rm dB$ \cite{2010Masada}, we see that photon subtraction applied on two modes lowers threshold squeezing parameter and enhances the fidelity in the practical squeezing regime. Threshold squeezing parameters for photon added CV GHZ states are $r_A=r_C\approx 0.477$, $r_B\approx 0.289$, $r_{AB}=r_{BC}\approx 0.443$, $r_{AC}\approx 0.426$, and $r_{ABC}\approx 0.484$, which are all larger than the value for the CV GHZ state.

\subsection{OPTIMAL GAIN}
One can adjust the gain $g$ to maximize the fidelity in each scheme. As is known, the gain $g=(e^{4r}-1)/(e^{4 r}+1/2)$ yields the optimal fidelity for the CV GHZ states, with the classical fidelity $1/2$ achieved even in a zero-squeezing limit $r\rightarrow0$ \cite{2002van_loock}. When maximized over $g$, the CV GHZ states with photon subtraction on two modes still achieves better fidelities over the CV GHZ states and it also approaches the classical fidelity $1/2$ as $r\to 0$. While the overall pictures of photon subtraction on one mode or on three modes show similar behavior to the unit gain cases, the fidelity of photon subtraction on three modes around $r\to 0$ is better than that of photon subtraction on one mode. Also photon subtraction on mode A or mode C around $r\to 0$ gives a value similar to that of photon subtraction on mode B. Threshold squeezing parameters for the optimized photon subtraction schemes are $r_A=r_C\approx 0.338$, $r_B\approx 0.264$, and $r_{ABC}\approx 0.384$, which are lower than those of the unit gain schemes.

Turning attention to photon added CV GHZ states with an optimized gain, the fidelity of photon addition on mode B notably becomes comparable to that of the CV GHZ state, reaching the classical best fidelity as $r\to 0$. All other cases become almost indistinguishable. Threshold squeezing parameters for optimized photon addition schemes are $r_A=r_C\approx 0.471$, $r_{AB}=r_{BC}\approx 0.436$, $r_{AC}\approx 0.422$, and $r_{ABC}\approx 0.450$, which are a little lower than those of the unit gain schemes.

\subsection{THREE-MODE EPR CORRELATIONS}
The multipartite EPR correlation among the quadratures in the form of Eq.~\eqref{eq:quadrature correlations} is closely related to the teleportation network protocol. Thus, it is meaningful to compare the behaviors of the EPR correlation and the teleportation fidelity for each considered state. Here we examine the sum of the quadrature correlations,
\begin{equation}
\langle[\Delta (x_i-x_j)]^2\rangle+\langle(\Delta\sum_i\nolimits p_i)^2,
\end{equation}
 as a function of the squeezing parameter $r$.
\begin{figure}[hbtp]
\centering
\includegraphics[width=0.4\textwidth]{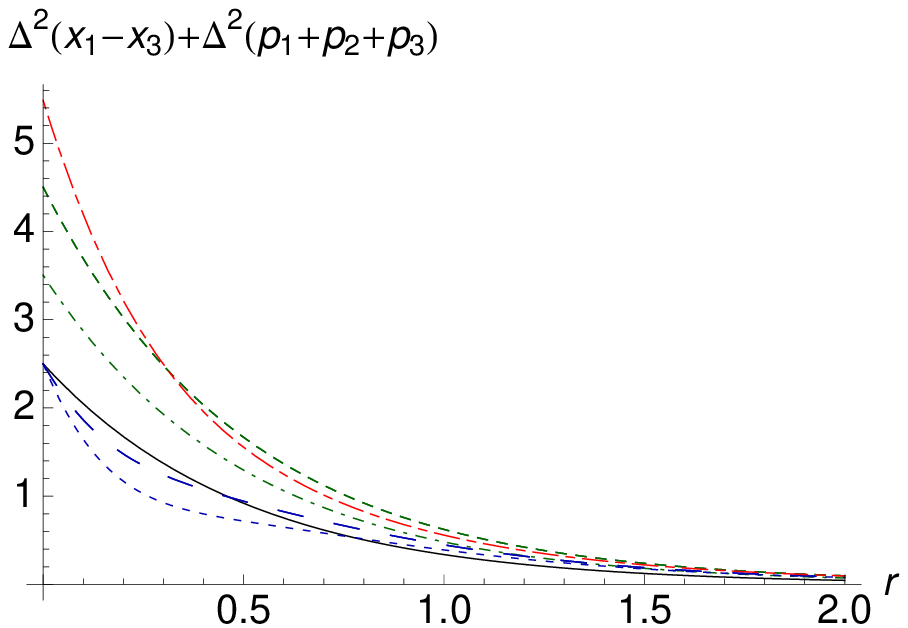}
\includegraphics[width=0.4\textwidth]{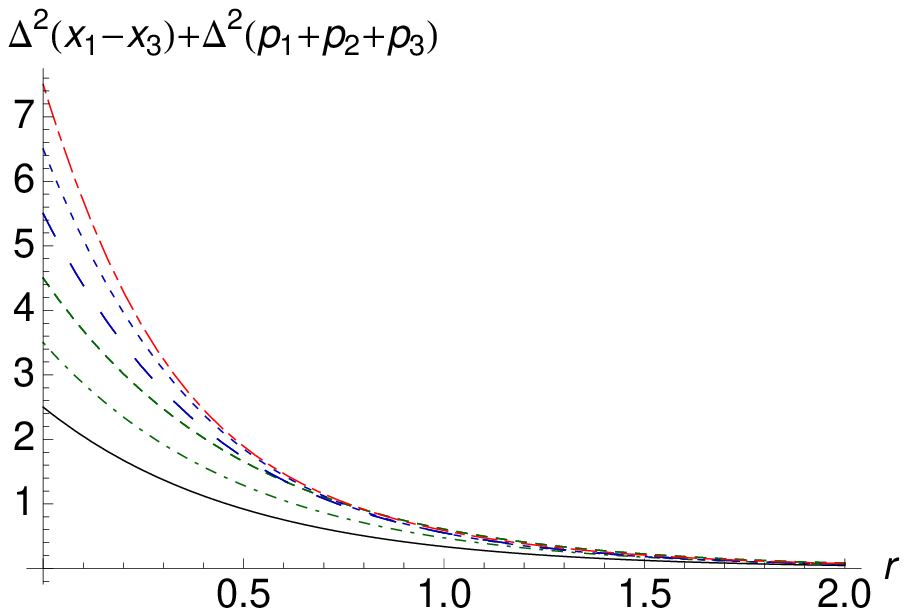}
\caption{(Color online) Quadrature correlations of the CV GHZ states with photon operations as a function of the squeezing parameter $r$. Top curves for photon subtraction and bottom curves for photon addition on mode A (green, small-dashed), mode B (green, dot-small-dashed), mode C (green, small-dashed), mode A and B (blue, long-dashed), mode A and C (blue, dotted), mode B and C (blue, long-dashed) and mode A, B and C (red, dot-long-dashed) in comparison to the CV GHZ state (black solid). The curves for photon operation on mode A and those on mode C coincide. The curves for photon operations on both modes A and B and those on modes B and C also coincide.}
\label{fig:quadrature correlations}
\end{figure}
In Fig.~\ref{fig:quadrature correlations} we see that photon subtraction on two modes again improves the quantum correlations between the quadratures in a weak squeezing regime similar to the case of the fidelity of the teleportation network protocol. However, details such as the crossing points of each graph are slightly different and this implies that the resources necessary for the teleportation network protocol are not exactly those quantum correlations expressed by the second moments of the quadratures. This was also noted for the case of two-mode CV teleportation in Ref. \cite{2012nha}. On the other hand, photon addition again turns out to  be unsuccessful in strengthen the tripartite quantum correlations.

\begin{figure}[tbh]
\centering
\begin{tabular}{cc}
\epsfig{file=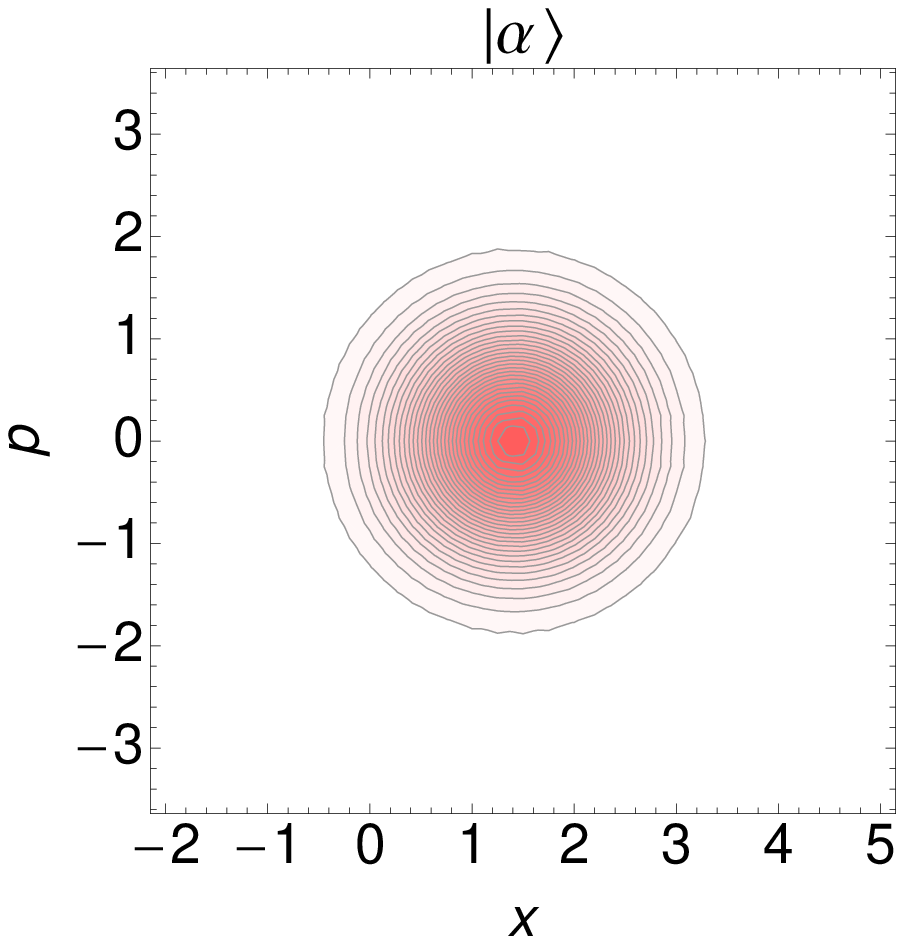,width=0.5\linewidth}&
\epsfig{file=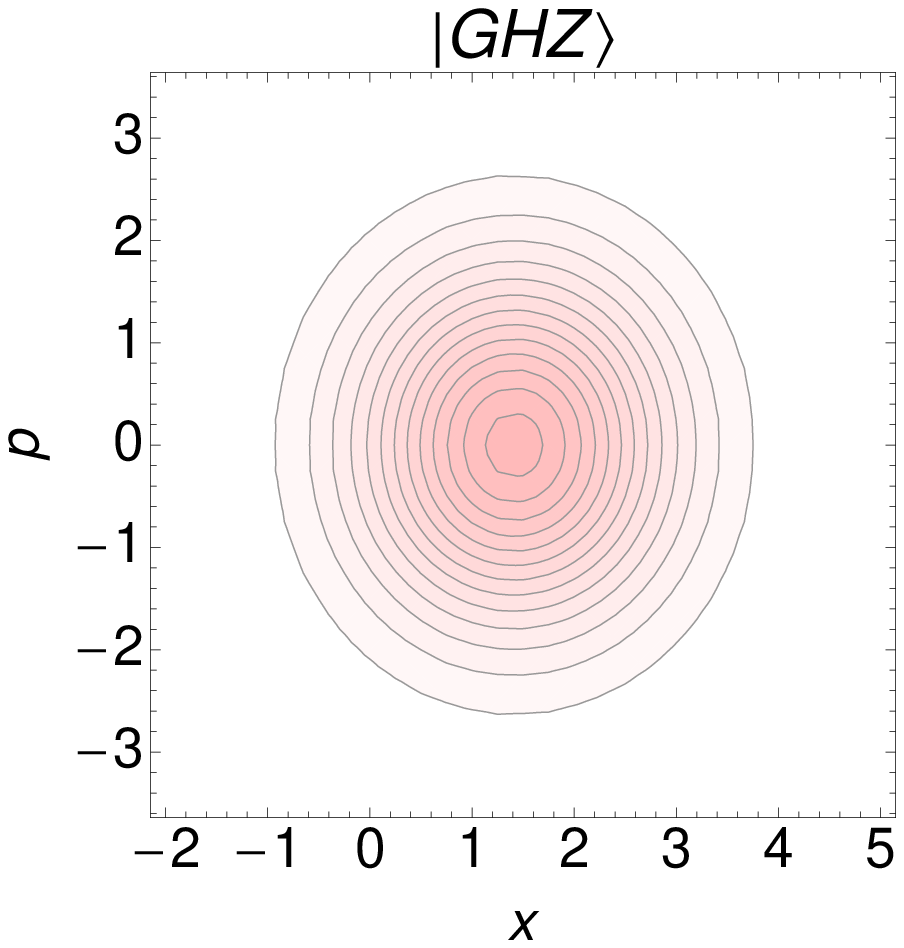,width=0.5\linewidth}
\end{tabular}
\caption{(Color online) Contour plots for input coherent state $\ket{\alpha=1}$ and the output state of the teleportation network protocol using the CV GHZ state (squeezing parameter $r=0.3$).}
\label{contour_coherent}
\end{figure}

\begin{figure}[tbh]
\centering
\begin{tabular}{cc}
\epsfig{file=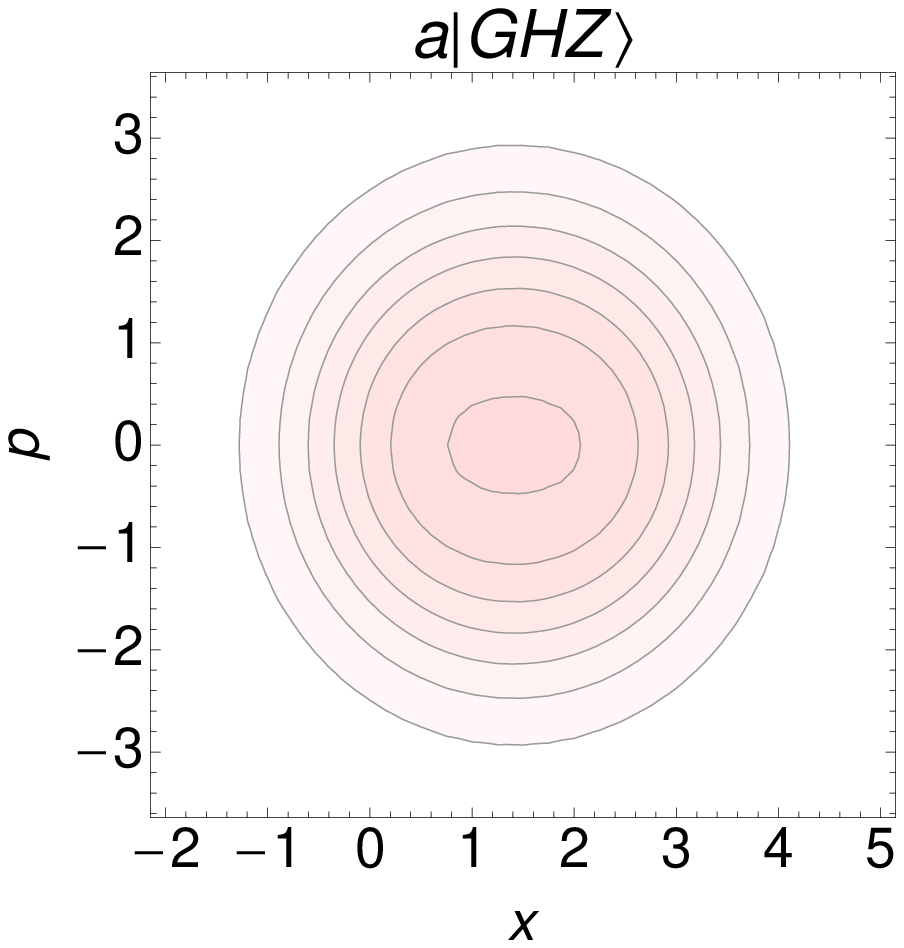,width=0.5\linewidth}&
\epsfig{file=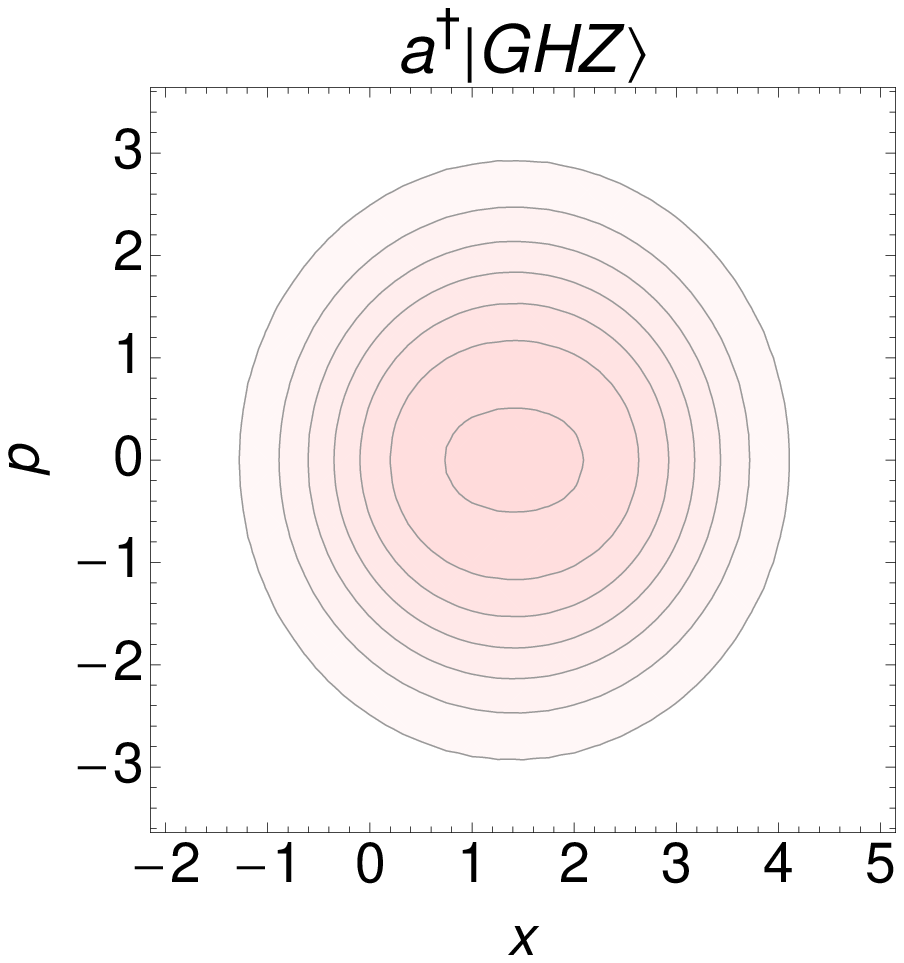,width=0.5\linewidth}\\
\epsfig{file=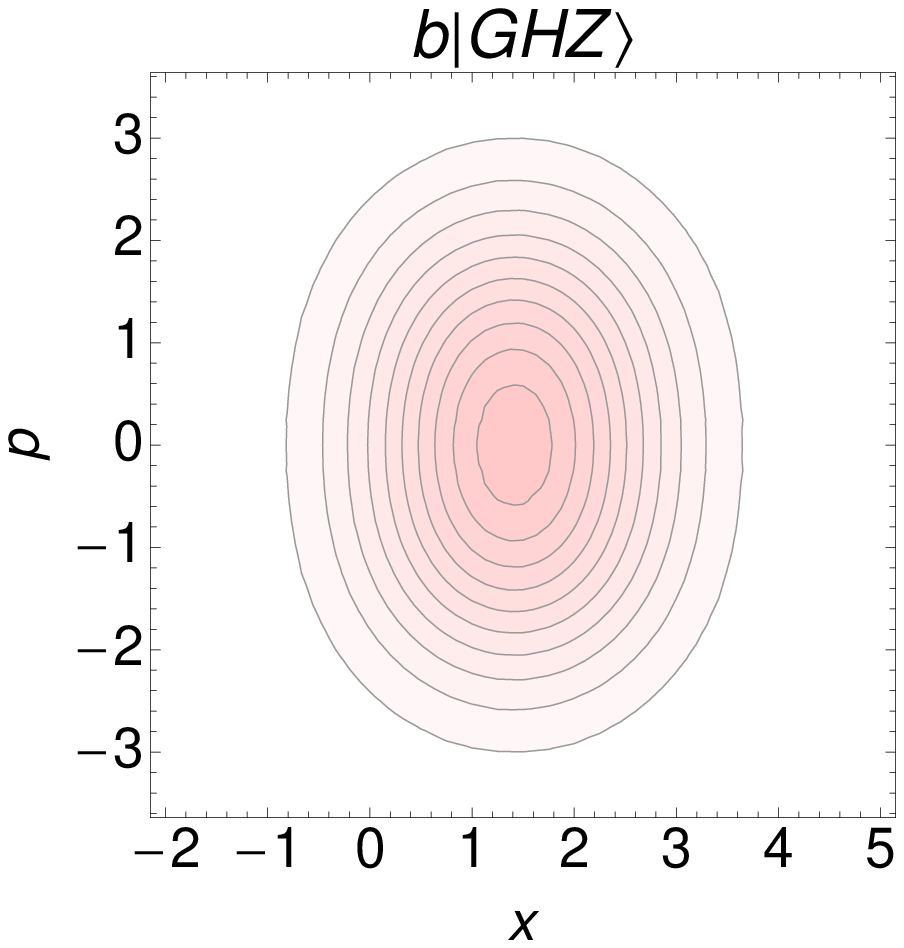,width=0.5\linewidth}&
\epsfig{file=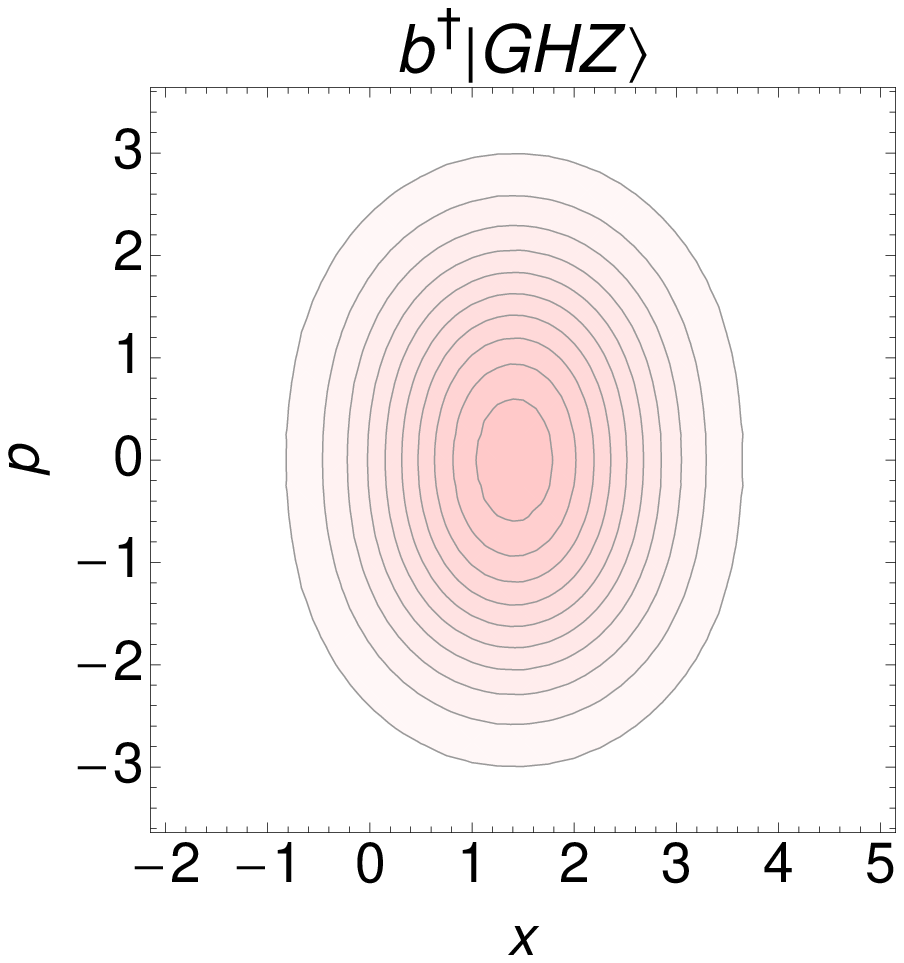,width=0.5\linewidth}\\
\epsfig{file=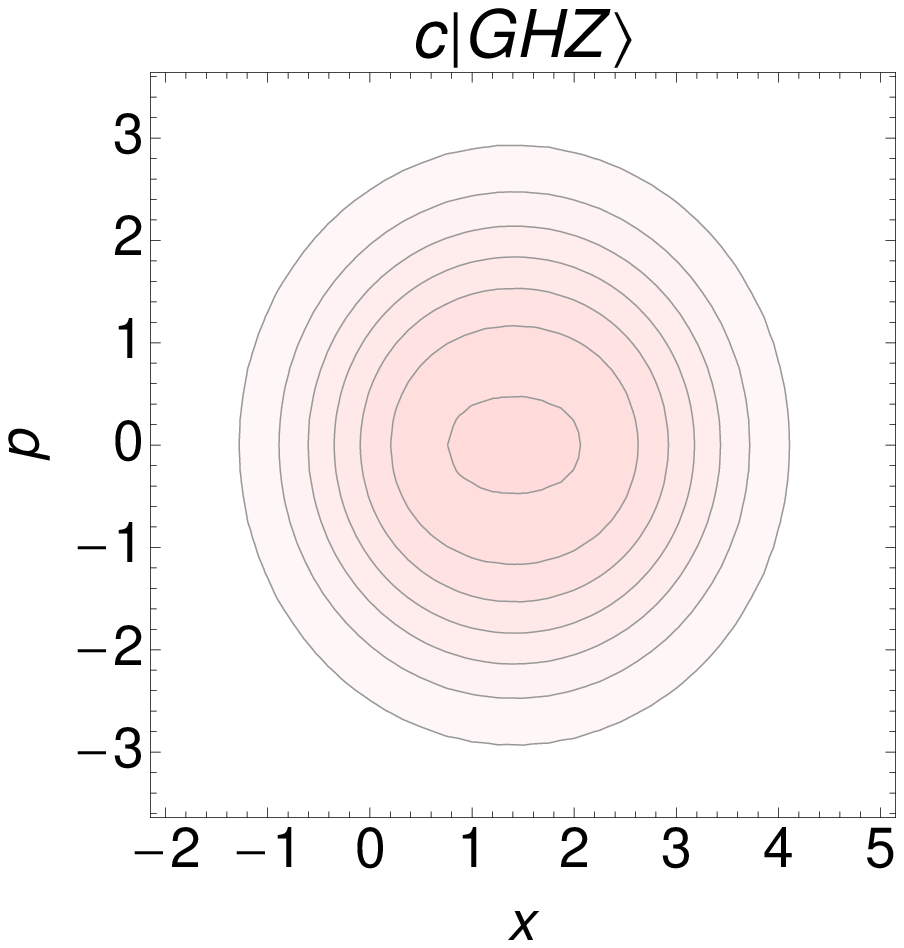,width=0.5\linewidth}&
\epsfig{file=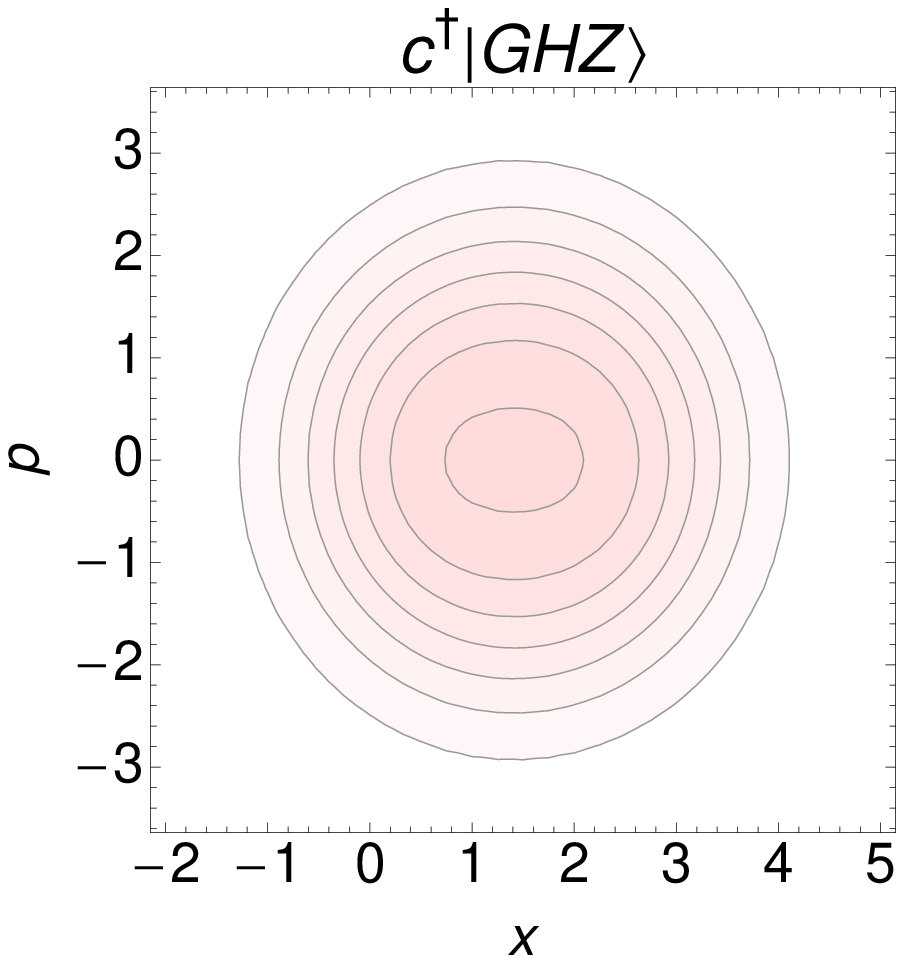,width=0.5\linewidth}
\end{tabular}
\caption{(Color online)) Contour plots for the output states of the teleportation network protocol using the CV GHZ state (squeezing parameter $r=0.3$) with photon operations on one mode.}
\label{contour_one_mode}
\end{figure}

\begin{figure}[tbh]
\centering
\begin{tabular}{cc}
\epsfig{file=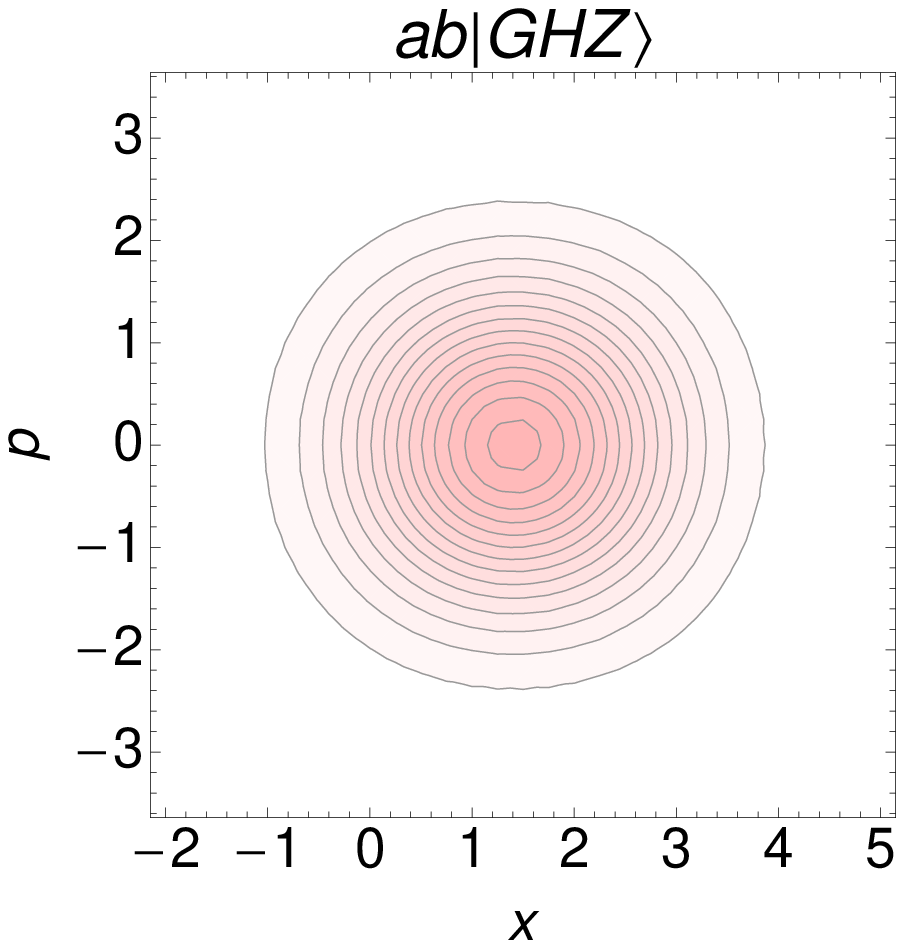,width=0.5\linewidth}&
\epsfig{file=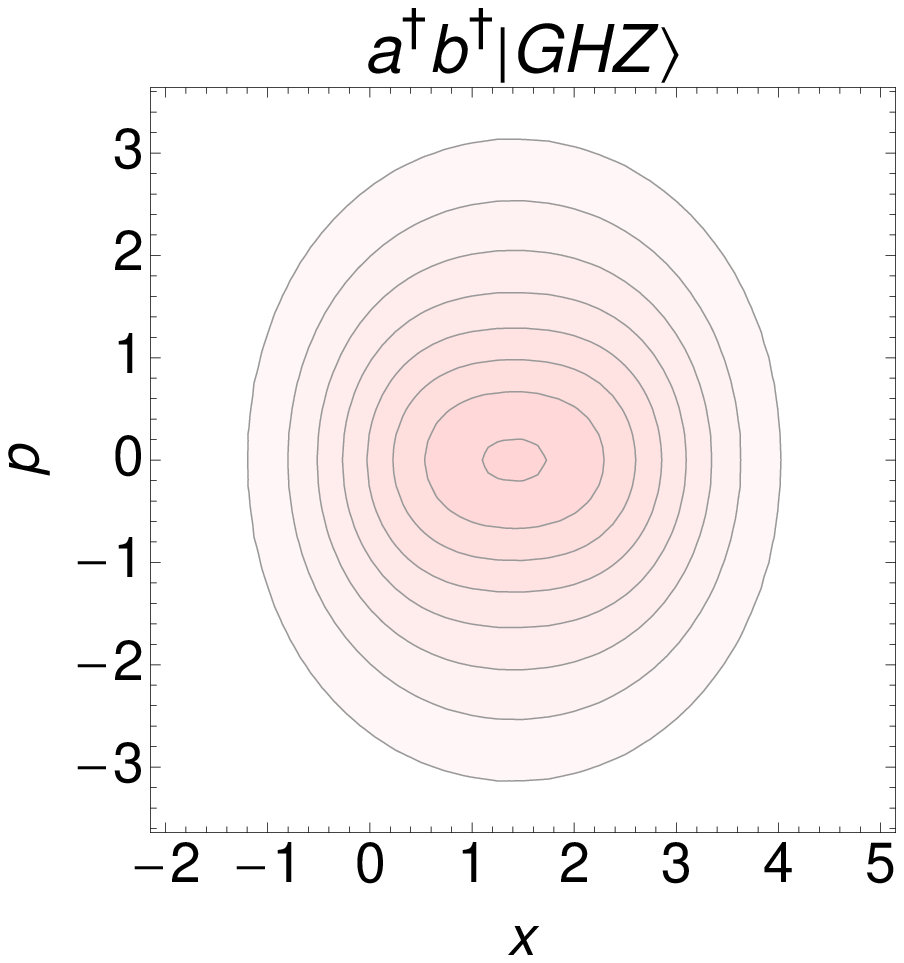,width=0.5\linewidth}\\
\epsfig{file=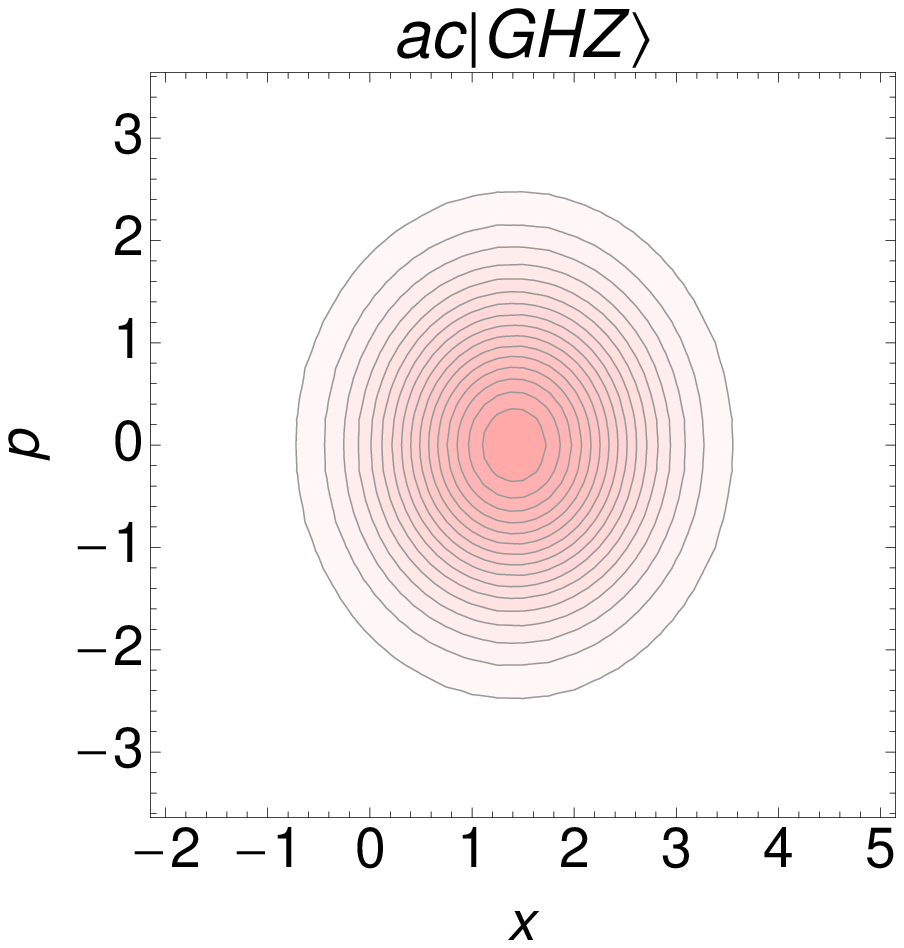,width=0.5\linewidth}&
\epsfig{file=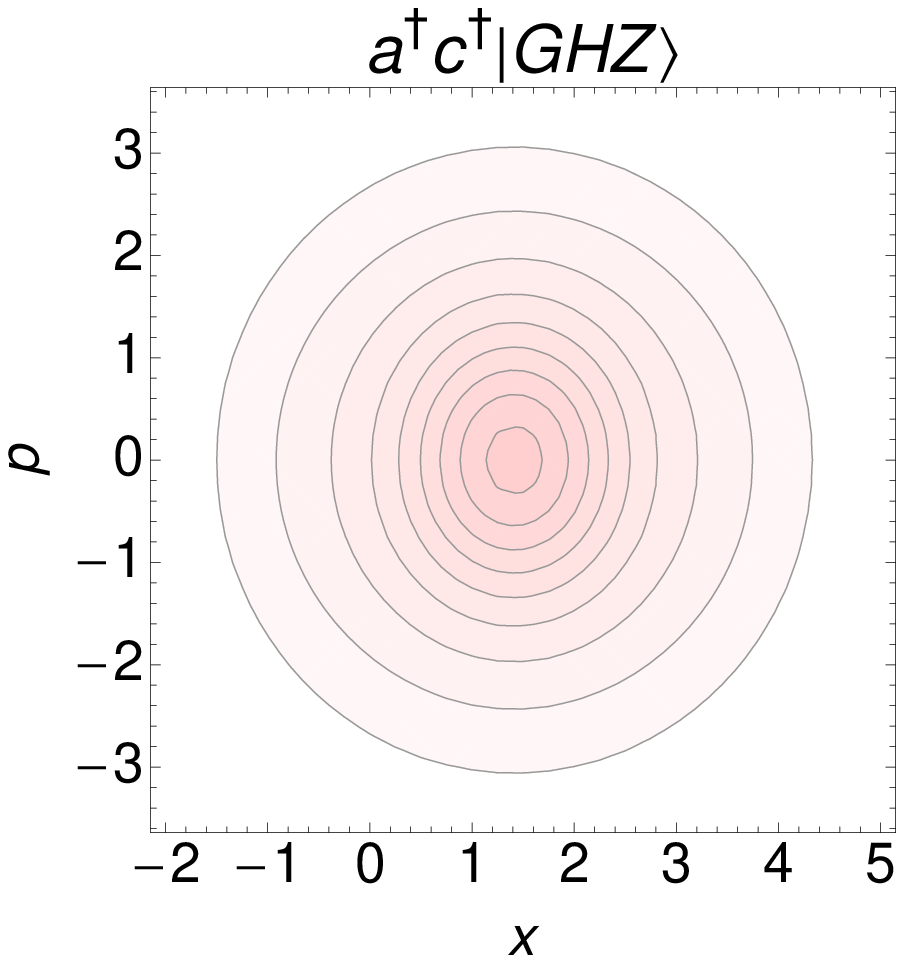,width=0.5\linewidth}\\
\epsfig{file=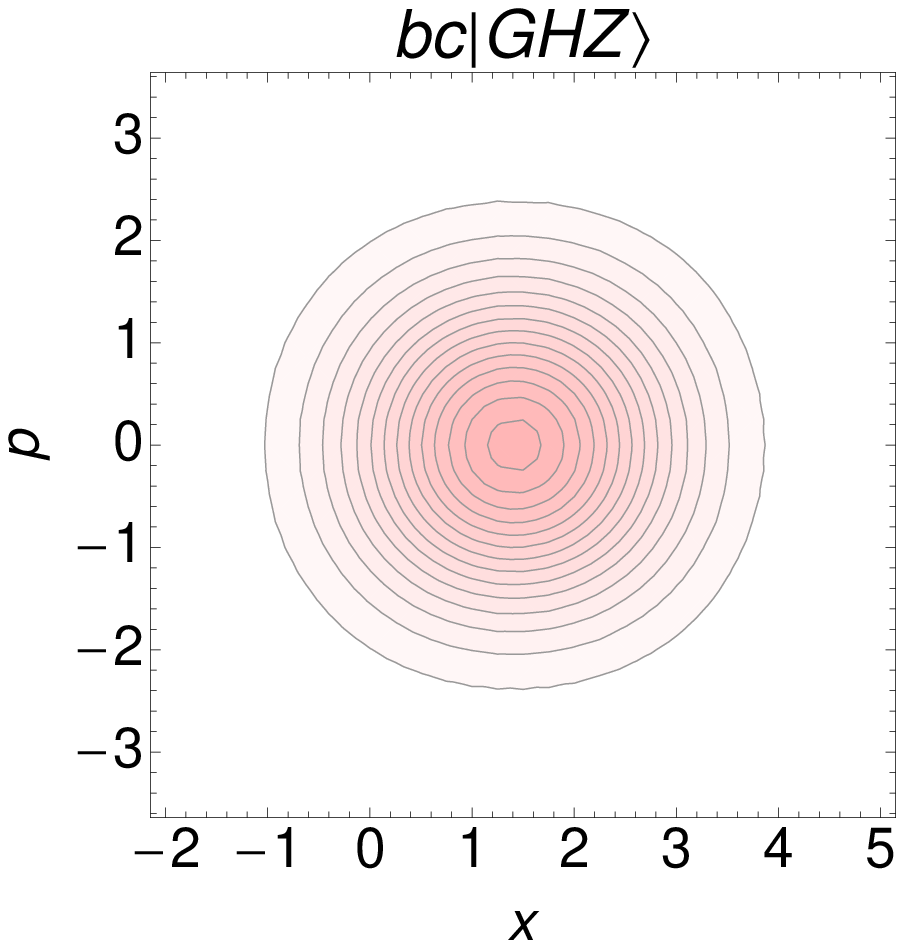,width=0.5\linewidth}&
\epsfig{file=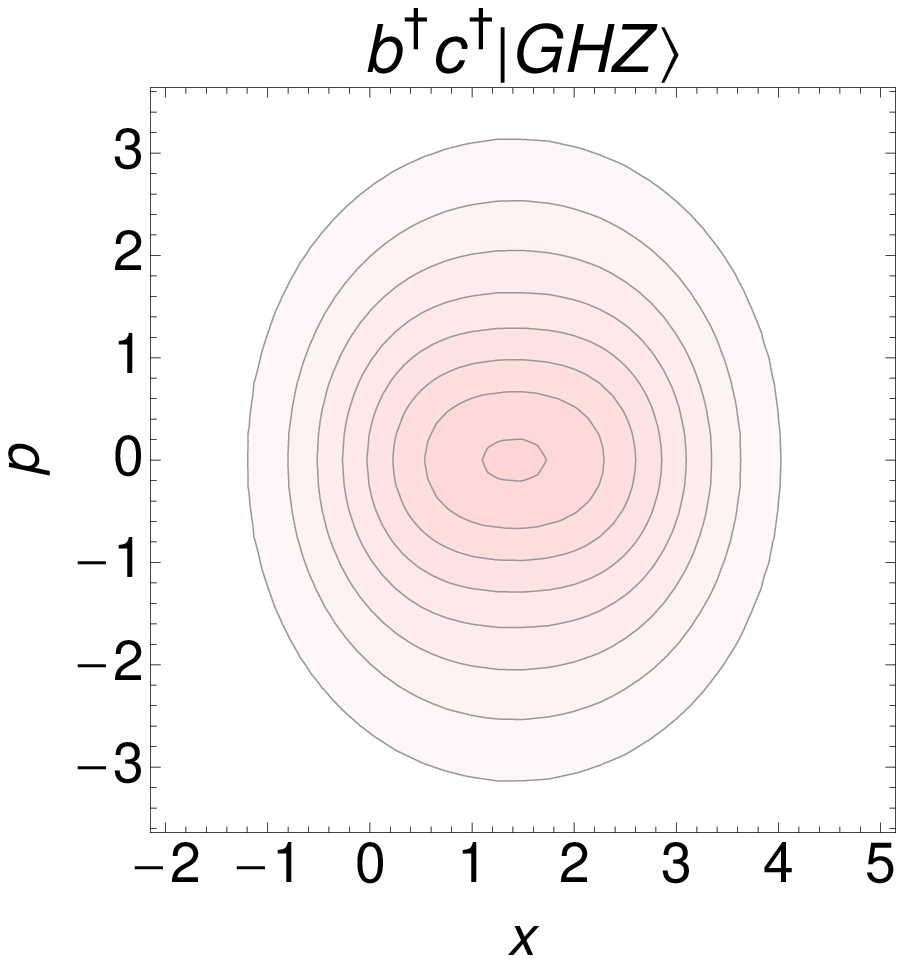,width=0.5\linewidth}
\end{tabular}
\caption{(Color online) Contour plots for the output states of the teleportation network protocol using the CV GHZ state (squeezing parameter $r=0.3$) with photon operations on two modes.}
\label{contour_two_modes}
\end{figure}

\begin{figure}[tbh]
\centering
\begin{tabular}{cc}
\epsfig{file=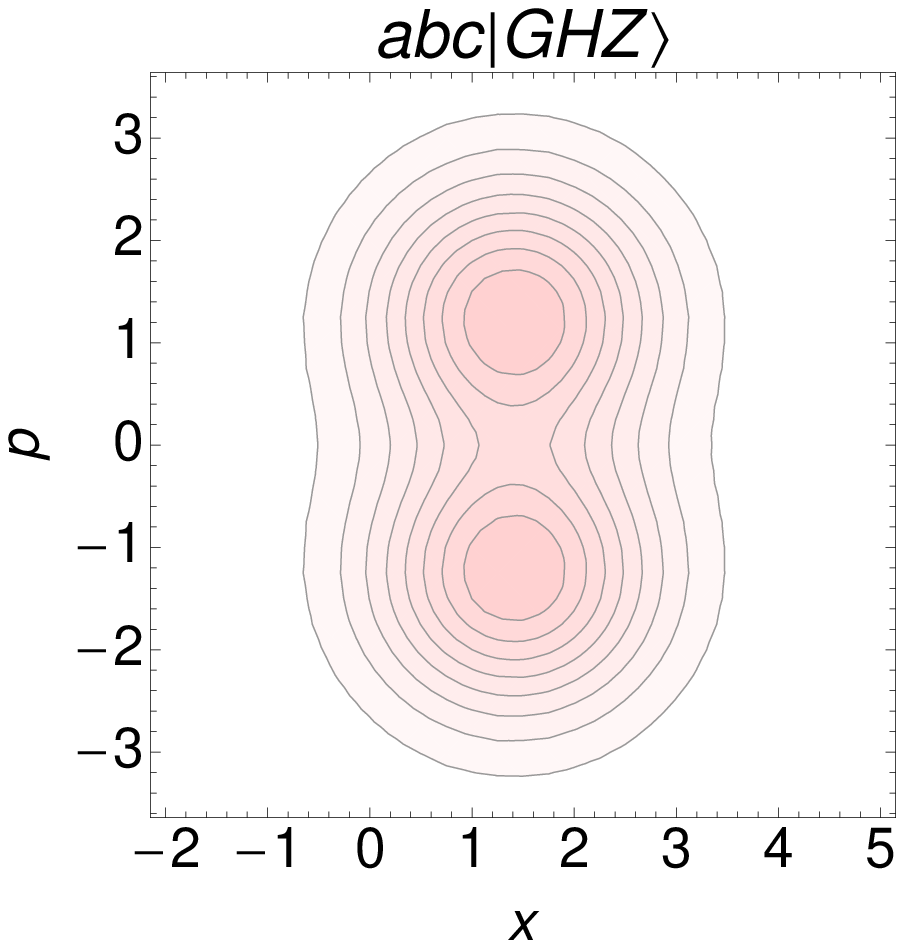,width=0.5\linewidth}&
\epsfig{file=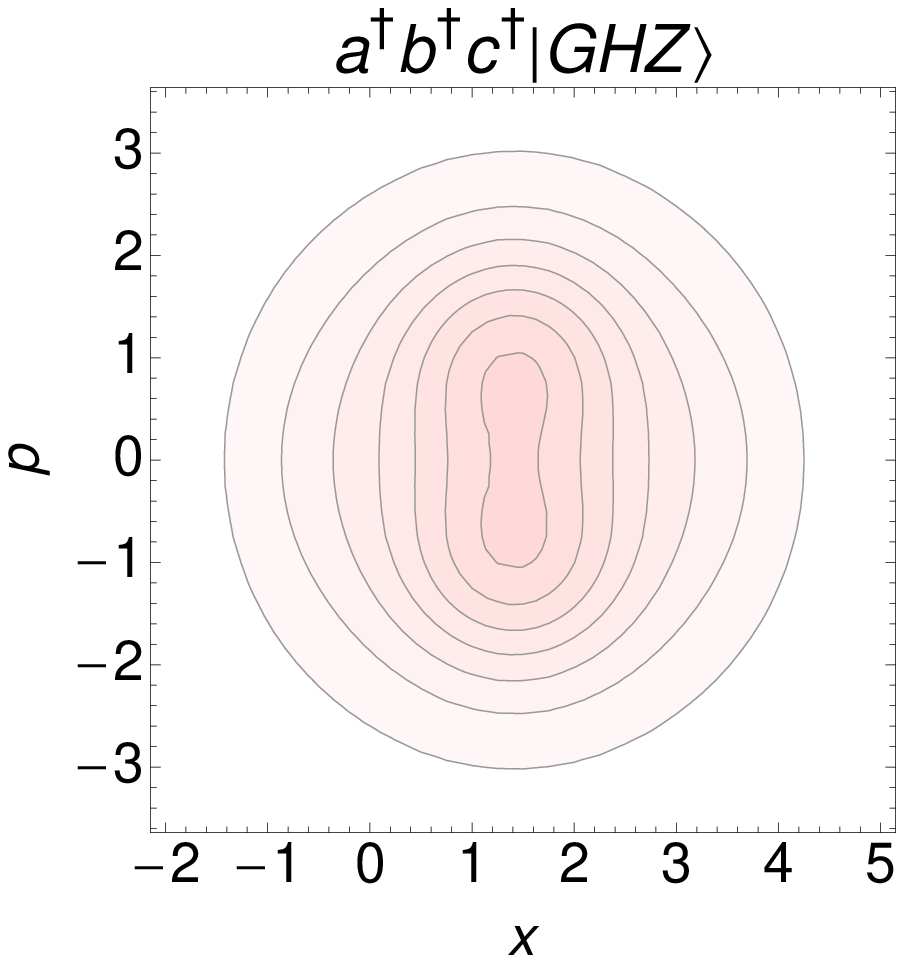,width=0.5\linewidth}
\end{tabular}
\caption{(Color online) Contour plots for the output states of the teleportation network protocol using the CV GHZ state (squeezing parameter $r=0.3$) with photon operations on three modes.}
\label{contour_three_modes}
\end{figure}

For further information, we examine contour plots of the Wigner distribution of the output states for each cases with unit gain $g=1$. Fig.~\ref{contour_coherent} shows the contour plots for the input coherent state $\ket{\alpha=1}$ and the output state of the teleportation network protocol using the CV GHZ state with a squeezing parameter $r=0.3$. Figs.~\ref{contour_one_mode}-\ref{contour_three_modes} give the contour plots of the output states using the CV GHZ state (squeezing parameter $r=0.3$) with photon operations applied on one mode, two modes, and three modes, respectively. In Fig.~\ref{contour_two_modes}, we see that photon subtraction on two modes keeps the spreading of the output state most tightly, hence giving the best result in terms of fidelity. Fig.~\ref{contour_three_modes} shows that photon operation on three modes pulls apart the distribution into two parts illustrating the worst fidelity they show.

\section{CONCLUSION}\label{sec:conclusion}
We have studied the characteristics of multipartite correlations for a class of non-Gaussian states which are created from the CV GHZ states applying the non-Gaussian photon operations. We have particularly investigated multipartite entanglement in terms of the Gaussian tangle, the multipartite nonlocality via MK inequality, and the fidelity of the teleportation network protocol together with the multipartite EPR correlations. Using three modes in our investigation, photon subtraction on two modes enhances the Gaussian tangle in a weak squeezing regime while other schemes do not show such an effect. As for the quantum nonlocality, photon subtraction generally enhances the violation of the MK inequalities in a weak squeezing regime. The maximum violation is given by photon subtraction on all three modes, which surpasses the maximum value achievable by the original CV GHZ state. Photon addition also enhances the violation in a broader regime of the squeezing parameter. Both the degree of violation for each squeezing parameter $r$ and the maximum values of the expectation value of the MK polynomial increase with the number of modes on which photon addition is applied, even one-mode-addition scheme showing maximum violation larger than that of the CV GHZ state. 

As for the fidelity of the teleportation network protocol, photon subtraction applied on two modes of sender and receiver shows an improved fidelity over that of the original protocol for a weak squeezing in both of the unit gain ($g=1$) and the optimized gain schemes. Among the one-mode subtraction schemes, photon subtraction on the mode of helper gives a higher fidelity than the other two cases. In contrast to the nonlocality and the two mode teleportation protocol, photon subtraction on all three modes shows the worst fidelity in a weak squeezing regime. On the other hand, photon addition fails to improve the fidelity for all the examined cases both in the unit gain ($g=1$) scheme and in the optimized gain scheme. Photon addition on the mode of helper gives the best fidelity among all photon-addition schemes for weak squeezing and its fidelity becomes comparable to that of the CV GHZ state when the optimized gain is used. With unit gain, photon addition on all three modes gives the worst fidelity in a weak squeezing regime, and all photon addition schemes except photon addition on the mode of helper give similar fidelities in the optimized scheme. The quadrature correlations expressed by a linear sum of covariances between quadratures also increase by photon subtraction on two modes similar to the fidelity of the teleportation network. 

Our results indicate that photon operations can be utilized to enhance various aspects of multipartite CV quantum correlations in a practical squeezing regime. 
There are still some interesting issues to further pursue, e.g. to find an optimal protocol for quantum communication using those non-Gaussian states created with photon operations or to validate the optimality of the original teleportation protocol.

\begin{acknowledgments}
Authors acknowledge useful discussions with Carlos Navarrete Benlloch and Se-Wan Ji. This work is supported by the NPRP grant 4-520-1-083 from Qatar National Research Fund and partly by the IT R\&D program of MOTIE/KEIT [10043464(2012)].
\end{acknowledgments}

\bibliography{reference_cv_GHZ_with_photon_op}

\end{document}